\documentclass[aps,prc,twocolumn,superscriptaddress,10pt,english,preprintnumbers, nofootinbib]{revtex4-2}
\pdfoutput=1
\usepackage[T1]{fontenc}
\usepackage[utf8]{inputenc}
\usepackage{epsfig}
\usepackage{graphicx}
\usepackage{xcolor}
\usepackage{latexsym}
\usepackage{textcomp}
\usepackage{amssymb}
\usepackage[colorlinks=true,linkcolor=blue,citecolor=blue, urlcolor=blue]{hyperref} 
\usepackage{amsfonts,amsthm,amstext,amscd}
\usepackage{amsmath}
\usepackage{mathtools}

\def\trento{T\raisebox{-0.5ex}{R}ENTo}
\def\zetamax{(\zeta/s)_{\rm max}}

\def\vtwo{$v_2\{2\}$}

\def\meanpt{$\langle p_T \rangle$}

\def\saa{\sigma_{AA}}
\def\snn{\sigma_{NN}}
\def\trento{T\raisebox{-0.5ex}{R}ENTo}
\def\dmin{$d_\text{min}$}
\def\sigmaf{$\sigma_\text{fluct}$}
\def\tauhyd{$\tau_\text{hyd}$}
\def\rhyd{$r_\text{hyd}$}
\def\taupipi{$\frac{\tau_{\pi\pi}}{\tau_\pi}$}
\def\chistruct{$\chi_\text{struct}$}

\DeclareMathOperator{\diag}{diag}

\begin{document}
\title{A generalized hydrodynamizing initial stage for Heavy Ion Collisions}%
\author{Govert Nijs}
\affiliation{Center for Theoretical Physics, Massachusetts Institute of Technology, Cambridge, MA 02139, USA}
\author{Wilke van der Schee}
\affiliation{Theoretical Physics Department, CERN, CH-1211 Gen\`eve 23, Switzerland}
\affiliation{Institute for Theoretical Physics, Utrecht University, 3584 CC Utrecht, The Netherlands}
\begin{abstract}
We present an extended Bayesian analysis using \emph{Trajectum} where the initial condition can now include binary scaling. For the far-from-equilibrium evolution before hydrodynamics we introduce an interpolation between free streaming and a holographically inspired evolution that exhibits fast hydrodynamization. We find strong evidence that binary scaling is incompatible with experimental data and find evidence that the holographic far-from-equilibrium evolution is preferred. We end with a discussion on several changes and improvements in the Bayesian framework.
\end{abstract}

\preprint{CERN-TH-2023-059//MIT-CTP/5547}

\maketitle

{\hypersetup{hidelinks}
\tableofcontents
}

\section{Introduction}
The main goal of heavy ion collisions (HIC) as performed at the Large Hadron Collider (LHC) in Geneva and the Relativistic Heavy Ion Collider (RHIC) in Brookhaven is to gain an understanding of the quark-gluon plasma (QGP) formed during such collisions. Such an understanding is however complicated by the fact that the several stages occurring during such a collision cannot be understood separately, but rather have to be understood together, as no experimental observable is sensitive exclusively to a single stage of the collision \cite{1301.2826,1802.04801}\@. After an initial interaction between the colliding nuclei far-from-equilibrium matter is formed. This matter eventually equilibrates, after which it can be described by relativistic hydrodynamics with a remarkably small shear viscosity \cite{0706.1522}\@. As the resulting hydrodynamic fluid expands and cools, it eventually reaches a temperature below which the fluid freezes out into particles, which eventually reach the various detectors.

Particularly poorly understood are the initial state and the pre-hydrodynamic stage. Some of the most widely used models for the initial state which are based on microscopic assumptions are AMPT \cite{nucl-th/0411110}, IP-Glasma \cite{1202.6646,1206.6805,1209.6330} and EKRT \cite{1505.02677,1310.3105, 2206.15207}, but for Bayesian analyses usually the phenomenological T\raisebox{-0.5ex}{R}ENTo model \cite{1412.4708} is chosen (see however the recent \cite{2302.09478} using IP-Glasma), which is followed by a free streaming stage for the pre-hydrodynamic stage (see however \cite{2302.14184} using anisotropic hydrodynamics)\@. The reason for this choice is that the T\raisebox{-0.5ex}{R}ENTo model has many parameters, allowing it to be tuned by Bayesian analysis to be compatible with experimental observables. The fitted parameters and their uncertainties can then be compared to behaviors from models such as IP-Glasma and EKRT, allowing the Bayesian analysis to act as a bridge between microscopic assumptions and experimental results, thereby giving evidence in favor of or against various microscopic models. As an example, \cite{1605.03954} showed that the scaling behaviors of the wounded nucleon model \cite{Bialas:1976ed,Shor:1988vk,Wang:1991hta,0805.4411,0710.5731} and the KLN model \cite{hep-ph/0111315,hep-ph/0212316,hep-ph/0408050} are incompatible with experimental data. More recently, the first inclusion of the PbPb hadronic cross section $\sigma_{AA}$ \cite{2204.10148manual} in a Bayesian analysis resulted in the conclusion that a drastically lower value for the nucleon size used in the T\raisebox{-0.5ex}{R}ENTo model is required for compatibility with the experimental value for $\sigma_{AA}$ \cite{2206.13522}\@.

In this work, we perform a Bayesian analysis using both an expanded initial state as well as a generalized pre-hydrodynamic stage. This Bayesian analysis is the same as the one used in \cite{2206.13522}, so in addition to the results presented here, this work also provides the details of the computation for \cite{2206.13522}\@.

We decided to first present our prime results, which are the posterior distribution for the generalized initial state and the far-from-equilibrium evolution. Only after these results we present the complete framework, including changes and improvements with respect to previous work. We end with a quantification of these changes and a discussion of all results.

\section{Generalized initial state}\label{sec:q}

A central part in any framework describing HIC is the initial condition at $\tau = 0^+$ or sometimes directly at the starting time of hydrodynamics \tauhyd\@. In early works this initial condition was often specified as a sum of `wounded nucleons' ($\mathcal{T}_A + \mathcal{T}_B$) and a `binary scaling' contribution ($\mathcal{T}_A\mathcal{T}_B$) \cite{Bialas:1976ed,Shor:1988vk,Wang:1991hta,0805.4411,0710.5731}\@. Here $\mathcal{T}_{A/B}$ is the thickness function of nucleus $A$ or $B$, which is given by the energy density in the transverse plane of nucleons that are part of the collision. At high energy or at weak coupling the binary scaling contribution is dominant, as the transparency of the collision dictates that all left- and right-moving nucleons interact with each other independently. At strong coupling a scenario more akin to full stopping arises, where all energy is put into the QGP \cite{1011.3562, 1305.4919}\@. This is similar to the Landau model \cite{Landau:1953gs} and more in line with the wounded nucleon scaling. In practice it was found that an 80\% contribution of wounded nucleons describes the centrality dependence of the multiplcity fairly well (see \cite{ALICE:2018tvkmanual} for a recent estimate of $f = 0.801$)\@.

Many recent models are more sophisticated than this simplified picture, and include microscopic insights from saturation physics \cite{2209.01176}, mini-jet deposition and saturation (EKRT, \cite{1505.02677,1310.3105}) or strong coupling \cite{1507.08195}\@. We will give more details on some of these models later, but continue here with a phenomenological prescription for the energy deposition called the T\raisebox{-0.5ex}{R}ENTo model \cite{1412.4708}\@. Here the energy density is given by
\begin{equation}
\mathcal{T} = N\left(\frac{\mathcal{T}_A^p + \mathcal{T}_B^p}{2}\right)^{1/p},\label{eq:trentoorig}
\end{equation}
with as parameters the norm $N$ and the T\raisebox{-0.5ex}{R}ENTo $p$ parameter. The motivation for this ansatz was that it is invariant under scaling ($\mathcal{T}\rightarrow c \mathcal{T}$ as $\mathcal{T}_{A/B}\rightarrow c \mathcal{T}_{A/B}$) and that it includes the wounded nucleon scaling as $p=1$\@. Previous Bayesian analyses have found $p \approx 0$ \cite{1605.03954,1808.02106,1904.08290,Bernhard:2019bmu,2010.03928,2011.01430,2010.15130,2010.15134,2110.13153}, in which case \eqref{eq:trentoorig} reduces to $\mathcal{T} = N\sqrt{\mathcal{T}_A\mathcal{T}_B}$\@.

This ansatz, however, is not based on a microscopic insight. As such, it can be used phenomenologically to compare with experiment and subsequently to compare with microscopic models to decide, for instance, if a model that resembles wounded nucleon scaling is ruled out by experimental data. This, however, is only possible so far as the ansatz can accurately describe the microscopic model, which is often far from clear. In particular, the ansatz cannot describe the binary scaling energy deposition and secondly it is \emph{a priori} unclear if the ansatz should describe the initial energy or entropy deposition. Indeed the first analysis using the \trento{} ansatz \cite{1605.03954} used it as an initial entropy deposition, whereas later analyses used it as an energy density. We will come back to this characteristic difference.

For these reasons it is desirable to extend the \trento{} ansatz to include a wider space of microscopic models. Here we do this by raising the ansatz to the power of an extra parameter $q$, according to
\begin{equation}
\mathcal{T} = NE_\text{ref}^{2-2q}\left(\frac{\mathcal{T}_A^p + \mathcal{T}_B^p}{2}\right)^{q/p},\label{eq:trentomodified}
\end{equation}
with $E_\text{ref}$ a constant with dimension of energy (see \cite{scott} for early results using the same ansatz)\@. It is important to note that while $E_\text{ref}$ is in principle redundant with the norm $N$, and as such does not have to be independently varied in Bayesian analyses, it is however important to choose a reasonable value for $E_\text{ref}$ if $q\neq 1$\@. The reason for this is that $E_\text{ref}$ sets an energy scale, which should be close to typical values of $\mathcal{T}_A$ and $\mathcal{T}_B$, so that
\[
\frac{1}{E_\text{ref}^2}\left(\frac{\mathcal{T}_A^p + \mathcal{T}_B^p}{2}\right)^{1/p} = \mathcal{O}(1).
\]
This in turn implies that when this number is raised to the power $q$, one does not obtain a very large or very small number. In this way, the correlation between $N$ and $q$ in the Bayesian analysis is minimized, which makes  smaller prior ranges for both of these parameters possible. In this work, we quantify this by choosing $E_\text{ref}$ so that the initial state entropy is reasonably independent of $q$. %
This results in a value of $E_\text{ref} = 0.2\,\text{GeV}$\@.

The generalized T\raisebox{-0.5ex}{R}ENTo formula given by \eqref{eq:trentomodified} now reduces to $\mathcal{T} \propto \mathcal{T}_A\mathcal{T}_B$ for $p = 0$, $q = 2$, and \eqref{eq:trentoorig} is recovered for $q = 1$\@. An additional advantage of \eqref{eq:trentomodified} is that now the model can approximately interpolate between interpreting $\mathcal{T}$ as an initial entropy or energy density. Early T\raisebox{-0.5ex}{R}ENTo studies interpret $\mathcal{T}$ as an entropy density $s$ \cite{1605.03954}\@. After the introduction of a free streaming pre-hydrodynamic stage, this was changed so that $\mathcal{T}$ was now interpreted as an energy density $\rho$ \cite{1808.02106}\@. Since the QCD equation of state is close to conformal for temperatures above the deconfinement transition, we approximately have $\rho \propto s^{4/3}$\@. This means that we can approximately go from interpreting $\mathcal{T}$ as an energy density to interpreting it as an entropy density by multiplying $q$ by $4/3$\@.

\subsection{Posterior distribution for $q$}

\begin{figure}[ht]
\includegraphics[width=0.7\columnwidth]{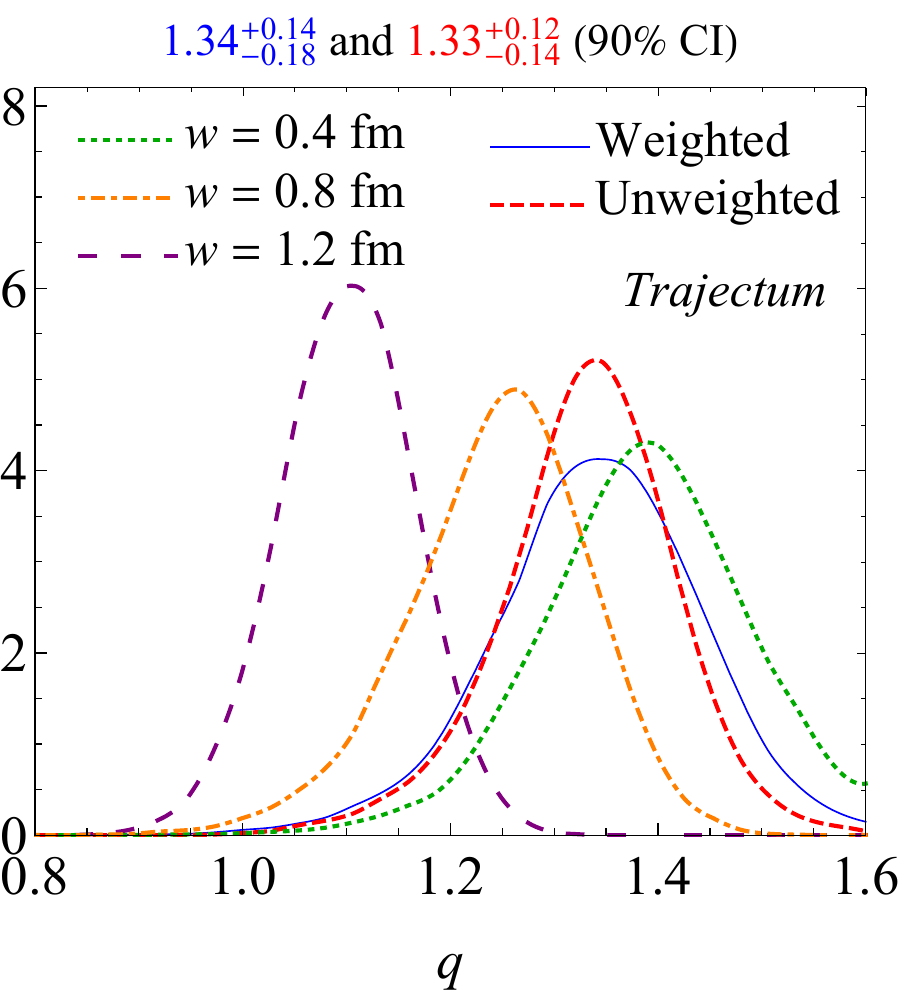}
\caption{\label{fig:qposterior}
We show the posterior distribution for the new $q$ parameter. For the full weighted result this gives $q=1.34^{+0.14}_{-0.18}$ at 90\% confidence. This value can be interpreted as scaling the original \trento{} ansatz (Eq.~\eqref{eq:trentoorig}) from an energy density to an entropy density. From the figure we see that $q$ depends strongly on the nucleon width and it is hence important that from \cite{2206.13522} we know that the nucleon width is relatively small at $w = 0.55^{+0.13}_{-0.14}\,$fm.
}
\end{figure}

In Fig.~\ref{fig:qposterior} we show the posterior distribution for $q$ for several fitting scenarios. The first is the full fit using all 653 data points, which is drawn as red dashed (full details are presented in Sec.~\ref{sec:trajectum})\@. A more realistic estimate underweights observables for which we trust the framework less, which in this case includes $p_T$-differential observables as well as particle identified observables. As expected this underweighting leads to less precise results, but the two distributions are in agreement. It turns out that the optimal $q$ value is highly correlated with the nucleon width $w$, which is clear from the three separate fits where we fix the nucleon width $w = 0.4$, 0.8 and $1.2\,$fm respectively. The variation of the width is an illustration of the importance of the total nuclear cross section measurement of $\saa{}$ for the determination of $q$, since it is mostly this measurement that implies a relatively small width \cite{2206.13522}\@.

It is interesting that $q$ peaks around $4/3$\@. Together with the narrow width of the posterior this strongly rules out an energy density that scales as binary scaling ($\mathcal{T}_A\mathcal{T}_B$), which would correspond to $q = 2$ \footnote{In principle a bimodal distribution for $q$ could exist, but using an early coarse analysis with larger prior ranges we could rule out such a scenario.}\@. A binary scaling for the entropy density (implying $q = 8/3$) is strongly excluded as well. Together with the fact that $p$ peaks around $p=0$ (see also later Fig.~\ref{fig:allposteriors}) we can conclude that the initial entropy density is consistent with $\sqrt{\mathcal{T}_A\mathcal{T}_B}$\@. This interpretation has the subtlety that at $\tau = 0^+$ the matter is not in thermal equilibrium and hence an entropy density is not strictly defined. A more conservative interpretation is hence that the energy density scales as $\rho \propto (\mathcal{T}_A\mathcal{T}_B)^{2/3}$\@. 

In future work it would be possible to consider analyses with $p = 0$, e.g.~of the form $\rho \propto (\mathcal{T}_A\mathcal{T}_B)^{q/2}$\@. This would have the advantage of limiting the number of parameters and keeping the model computationally tractable, though we note that there is no strong physical reason for such a form and hence it is possible that this artificially constrains the model too much to be realistic.

\subsection{Energy deposition in holography}

In the next three subsections we explore the energy deposition in three microscopic models in order to compare with the $\rho \propto (\mathcal{T}_A\mathcal{T}_B)^{q/2}$ scaling as just presented. The first is a holographic model where lumps of strongly coupled matter collide. Even though this is strictly speaking only valid at infinitely strong coupling we will argue that most results apply in any theory that is approximately scale invariant.
Secondly we will review analytic estimates from  color glass condensate (CGC) effective theory as well as numerical estimates from IP-Glasma simulations. These can be viewed as more weakly coupled results. These results differ, and one important reason for the difference is the extra scale provided by the saturation scale $Q_{s,0} = 0.794\,$GeV\@.

\begin{figure*}[ht]
\includegraphics[width=\textwidth]{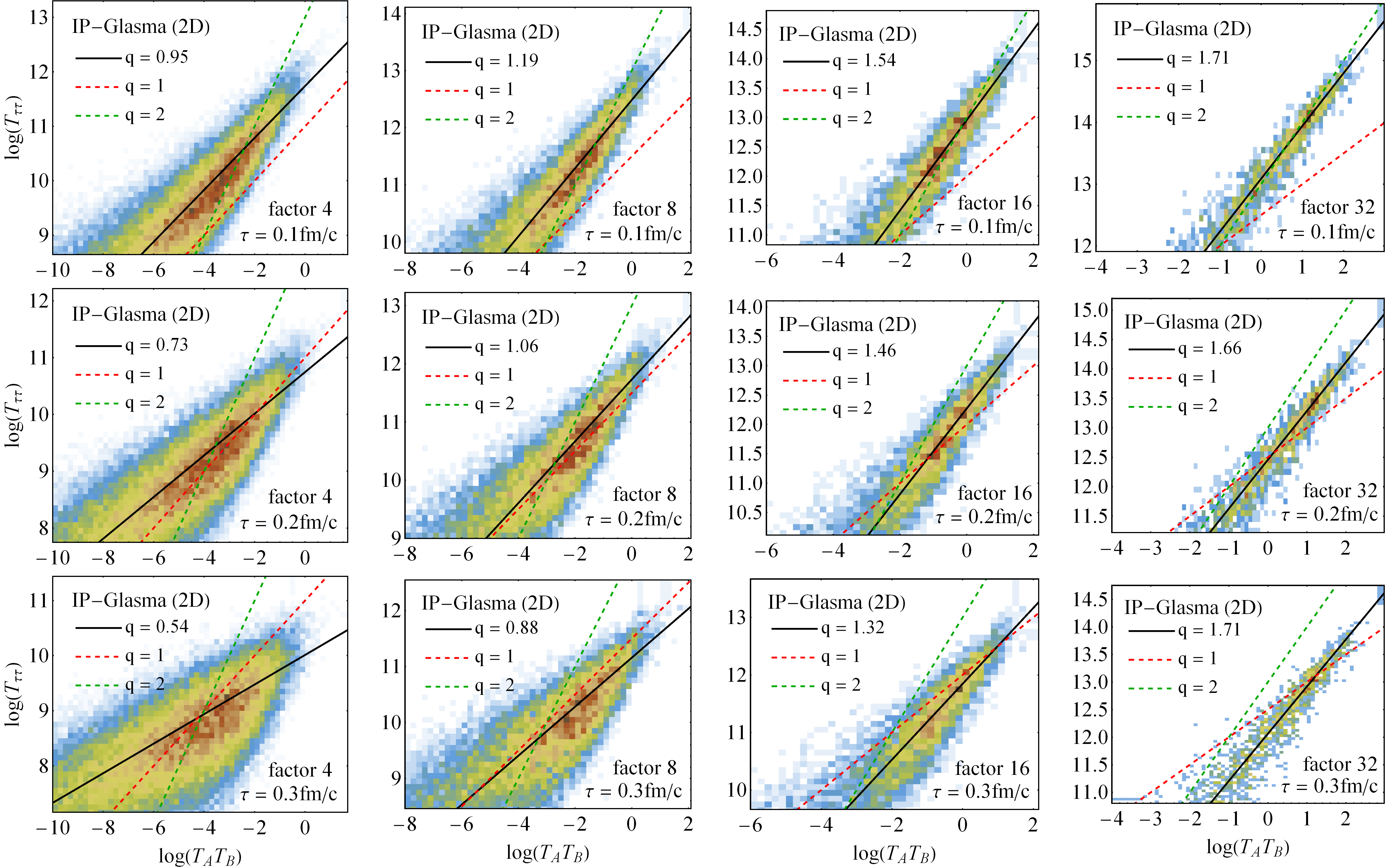}
\caption{\label{fig:ipglasma}We show histograms of the logarithm of the energy density $T_{\tau\tau}$ versus the logarithm of the product of the thickness functions $\mathcal{T}_A \mathcal{T}_B$ averaged over 100 IP-Glasma evolutions at times 0.1, 0.2 and 0.3 fm (three rows) and coarse-grained by factors 4, 8, 16 and 32 (four columns) for PbPb collisions at 5.02 TeV\@. The lines illustrate the $\rho \propto (\mathcal{T}_A\mathcal{T}_B)^{q/2}$ scaling for $q$ of 1, 2 and a best fit. The best fit values increase with coarse-graining and generically decrease with time. Especially after coarse-graining by a factor 16 (corresponding to a lattice spacing of $0.94\,$fm) the results are in general agreement with the posterior of Fig.~\ref{fig:qposterior}\@.}
\end{figure*}

In holography the energy density $\rho$ at mid-rapidity immediately after the collision of two shocks of energy scales as \cite{0803.3226, 1305.4919, 1307.2539, 1712.05815}:
\begin{equation}
\rho \propto \mathcal{T}_A \mathcal{T}_B \tau^2 + \mathcal{O}(\tau^5),\label{eq:holography}
\end{equation}
with $\tau$ the proper time.
While this relation, including the normalization factors, can be rigorously derived in holography, we note that the binary scaling proportionality basically follows from linearity close to the holographic boundary (this is valid for short times) and the $\tau^2$ then follows from dimensional analysis. It is hereby important that at least in the simplest holographic setting the theory is both scale and Lorentz invariant, which means that the only relevant energy scale is given by $\mu = (\mathcal{T}_A \mathcal{T}_B)^{1/6}$\@.

The very first dynamics is approximately boost invariant \cite{0803.3226} and it may hence seem counter-intuitive that the energy density is increasing, instead of the usual $1/\tau$ decrease one would have in a boost invariant expanding plasma of weakly interacting particles. The energy for this increase is coming from the delta-like shocks of energy, but this increase indeed cannot last forever. The time where the maximum energy density is reached can also be derived from dimensional analysis and is proportional to $\tau_\text{max} \propto 1/\mu$\@. After this maximum is reached the system can quickly be described by relativistic hydrodynamics (it hydrodynamizes), which is again fixed by dimensional analysis:
\begin{equation}
\rho \propto \mu^{8/3} \tau^{-1 - 1/3} + \mathcal{O}(\tau^{-2}).\label{eq:latetime}
\end{equation}
Here the $-1/3$ is due to the work performed by the pressure, which we assumed to be $\mathcal{P}=\rho/3$\@. We note that in the \trento{} formula this form corresponds to $p=0$ and $q=8/9$\@.

It is important here that the former argument may have come from holography, but in the end the argument is completely fixed by scale invariance and dimensional analysis. As such it is not necessarily an argument that necessitates strong coupling. Moreover, it is known that a simple scale invariant estimate is not sufficient to describe heavy ion collisions. This can already be seen from the dependence of the multiplicity versus the beam energy \cite{0805.1551, 0902.1508}\@. By dimensionality it is there derived that $N_{\rm ch} \propto \left(\sqrt{s_{\rm NN}}\right)^{2/3}$, whereas the experimental scaling is closer to $\left(\sqrt{s_{\rm NN}}\right)^{0.3}\log\left(s_{\rm NN}\right)$ than $\left(\sqrt{s_{\rm NN}}\right)^{2/3}$ \cite{1304.0347, 1612.08966}\@. This hence implies a need for an extra scale \cite{1111.1931}, much like the QCD confinement scale $\Lambda_{\rm QCD}$, which will change both this multiplicity dependence as well as the simplified formula \eqref{eq:latetime}.

\subsection{Energy deposition in CGC}

Alternatively it is possible to study the energy deposition at weak coupling. Since this is dominated by low momentum gluons the appropriate description is then color glass condensate (CGC) effective field theory. In this case the relevant scale is not set by the colliding energy, but by the colliding (color) charge. This has hence a dimension $1/[\text{length}]^2$ instead of $[\text{energy}]/[\text{length}]^2$, and it is therefore no surprise that the energy deposition is initially constant: 
\begin{equation}
\rho \propto n_A n_B + \mathcal{O}(t^2)\label{eq:cgc},
\end{equation}
where now as mentioned $n_A$ refers to the charge density instead of the energy density.

After some time this energy density becomes proportional to \cite{2209.01176}
\begin{equation}
\rho \propto \frac{\mathcal{T}_A \mathcal{T}_B (2 \mathcal{T}_A^2 +  7\mathcal{T}_A \mathcal{T}_B  +2\mathcal{T}_B^2)}{(\mathcal{T}_A + \mathcal{T}_B)^{5/2}} \label{eq:cgc},
\end{equation}
which is approximately proportional to $(\mathcal{T}_A \mathcal{T}_B)^{3/4}$ if $\mathcal{T}_A$ and $\mathcal{T}_B$ are approximately equal. This corresponds to $p=0$ and $q = 3/2$, which is compatible with our posterior.

\subsection{Energy deposition in IP-Glasma}

Of special interest is the energy deposition in the phenomenologically successful IP-Glasma model \cite{1202.6646,1206.6805,1209.6330}\@. This model uses an impact parameter (IP) dependent saturation scale (IP-sat, \cite{hep-ph/0304189}) and subsequently uses this to evolve the classical Yang-Mills equations of the CGC to construct an initial state for hydrodynamics. The code is publicly available \cite{ipglasmagithub} and in this work we generate 100 such evolutions using standard settings appropriate for PbPb collisions at $5.02\,$TeV\@. It is important to note that the publicly available code is a boost invariant 2D code. This 2D analysis may not be equivalent to an approximately boost invariant 3D evolution that is restricted to mid-rapidity (see \cite{1605.07158, 1703.00017, 2001.08636, 2010.11172})\@. It would hence be interesting to repeat this analysis with a 3D code (when publicly available) and then restrict those to mid-rapidity.

An important feature of IP-Glasma is that it has finer structure than e.g.~the \trento{} model (see \cite{1206.6805} for an illustration)\@. This requires a fine lattice spacing and hence also makes the computations slightly expensive (we output at a resolution of $0.06\,$fm)\@. After some time, however, the plasma is expected to become smoother and can be transferred to a hydrodynamic code. For this purpose it is hence important to understand coarse grained energy deposition in the IP-Glasma model.

In Fig.~\ref{fig:ipglasma} we present histograms of the logarithm of the energy density $T_{\tau\tau}$ versus the logarithm of the product of the thickness functions $\mathcal{T}_A \mathcal{T}_B$\@. Importantly we do this for three different times (three rows at 0.1, 0.2 and 0.3 fm$/c$ respectively) and at a coarse grained level with factors 4, 8, 16 and 32 (four columns)\@. 

From Fig.~\ref{fig:ipglasma} it is clear that IP-Glasma does not yield a unique $\rho \propto (\mathcal{T}_A\mathcal{T}_B)^{q/2}$ scaling. Instead, coarse-graining regions of 0.23 and 0.46 fm (corresponding to factors 4 and 8) result in a relatively wide distribution of $\rho$ versus $\mathcal{T}_A \mathcal{T}_B$, especially at later times. The more coarse grained histograms have a higher correlation and it can be seen that the best-fit $q$ values decrease with time and increase with coarse-graining. Especially the factor 16 coarse-grained results yield to $q$ values that are consistent with the posterior in Fig.~\ref{fig:qposterior}, which can partially explain the phenomenological success of the IP-Glasma model.

\section{An initial stage between free streaming and strong coupling}\label{sec:rhydro}

With the initial condition starts a far-from-equilibrium evolution until the time when hydrodynamics applies (the hydrodynamization time \cite{1610.02023})\@. With the \trento{} initial condition so far all studies have modeled this stage using a free streaming approach (with unit velocity \cite{Bernhard:2019bmu, 2011.01430} or an effective velocity $v_\text{fs}$ \cite{2010.15134})\@. This could be called a weakly coupled approach, since at weak coupling particles do not interact much and hence free stream. The transition to hydrodynamics, on the other hand, is not smooth, since the hydrodynamic evolution is characteristically different from free streaming evolution. As such, even at weak coupling one would need a more complete model of hydrodynamization to realistically capture the far-from-equilibrium initial stage \cite{1506.06647, 1805.01604}\@. 

It is an interesting question if such a weakly coupled approach to the pre-hydrodynamic stage of a heavy ion collision would work. After all, just after the hydrodynamization time $\tau_\text{hyd}$ the collision can be described well by a hydrodynamic fluid with small shear viscosity, implicating the fluid is strongly coupled at that time. In this work we will hence supplement the free streaming approach with a holographically inspired approach. In holography the far-from-equilibrium matter hydrodynamizes quickly, meaning that within a time $1/T$, with $T$ the temperature at that time, the fluid obeys the first order viscous hydrodynamic equations \cite{0812.2053, 1011.3562, 1103.3452, 1202.0981, 1703.09681}\@.

For free streaming assuming boost invariance the full stress tensor at \tauhyd{} is given by
\begin{align}
T^{\mu\nu}(x,y) & = \frac{1}{2\pi\tau_\text{hyd}}\int_0^{2\pi}d\phi\,\hat p^\mu\hat p^\nu \nonumber\\
& \times \mathcal{T}(x - \tau_\text{hyd}\cos\phi, y - \tau_\text{hyd}\sin\phi),\label{eq:free}
\end{align}
with $\tau_\text{hyd}$ the hydrodynamization time and
\[
\hat p^\mu\hat p^\nu = \left(\begin{array}{ccc}
1 & \cos\phi & \sin\phi \\
\cos\phi & \cos^2\phi & \cos\phi\sin\phi \\
\sin\phi & \cos\phi\sin\phi & \sin^2\phi
\end{array}\right),
\]
where $\mathcal{T}$ is then given by the energy density at $\tau=0^+$
(see Eq.~\eqref{eq:trentoorig}) \cite{1808.02106}\@. This stress tensor is then decomposed \footnote{The decomposition can be explicitly done by using that $\Delta^{\mu\nu}$ and $\pi^{\mu\nu}$ are both orthogonal to $u^\mu$ and by using that $\pi^{\mu\nu}$ is traceless. The precise procedure is discussed in detail in Sec.~3.2.1 of \cite{1804.06469}\@.} as
\begin{equation}
T^{\mu\nu} = \rho u^\mu u^\nu - (P + \Pi)\Delta^{\mu\nu} + \pi^{\mu\nu},\label{eq:decomp}
\end{equation}
with $\Delta^{\mu\nu} \equiv g^{\mu\nu} - u^\mu u^\nu$, $g_{\mu\nu} = \diag(1, -1, -1, -\tau^2)$, $u^\mu$ the fluid velocity, $\rho$ the energy density, $P$ the pressure (given by the equation of state), $\Pi$ the bulk viscous pressure and $\pi^{\mu\nu}$ the shear tensor. 

Curiously, both in free streaming and in holographic theories for small gradients and small times the fluid velocity in the transverse directions $u_i$ is given by
\begin{equation}
  u_i = -\tfrac{1}{3} \tau_\text{hyd} \nabla_i \log(\mathcal{T}) +   \mathcal{O}(\tau_\text{hyd}^2).\label{eq:preflow}
\end{equation}
For free streaming (Eq.~\ref{eq:free}) the 1/3 is exact, but in holography the 1/3 is approximate \cite{1409.0040} and based on numerical holographic simulations performed in \cite{1211.2218, 1307.2539}\@. In hindsight this agreement is not very surprising. The development of flow in the transverse plane was studied precisely in \cite{0810.4325} and only depends on the pressure anisotropy between the longitudinal and transverse directions. For free streaming the longitudinal pressure vanishes because of boost invariance. In the holographic computations the longitudinal pressure starts out very negative (at minus two times the energy density, see also \cite{0803.3226, 1305.4919}), but becomes positive quickly due to the fast hydrodynamization. Dependening somewhat on time this longitudinal pressure averages out to zero, which explains why Eq.~\ref{eq:preflow} agrees for free streaming and holography. It is an interesting question how this works in weakly coupled models such as \cite{1805.01604}\@.

In a way this agreement is unfortunate, since $u_i$ has a strong effect on experimental observables, and as such it cannot distinguish between free streaming and holography. As we will see this is different for the shear tensor and bulk viscous pressure.

In hydrodynamics, the stress tensor is evolved according to second order viscous hydrodynamics in the 14 moment approximation \cite{1403.0962}, given by
\begin{align}
\partial_\mu T^{\mu\nu} & = 0,\nonumber \\
-\tau_\Pi D\Pi & = \Pi + \zeta\nabla\cdot u\label{eq:bulk} \\
& \quad + \delta_{\Pi\Pi}\nabla\cdot u\Pi - \lambda_{\Pi\pi}\pi^{\mu\nu}\sigma_{\mu\nu},\nonumber \\
-\tau_\pi\Delta_\alpha^\mu\Delta_\beta^\nu D\pi^{\alpha\beta} & = \pi^{\mu\nu} - 2\eta\sigma^{\mu\nu}\label{eq:shear} \\
& \quad + \delta_{\pi\pi}\pi^{\mu\nu}\nabla\cdot u - \phi_7\pi_\alpha^{\langle\mu}\pi^{\nu\rangle\alpha}\nonumber \\
& \quad + \tau_{\pi\pi}\pi_\alpha^{\langle\mu}\sigma^{\nu\rangle\alpha} - \lambda_{\pi\Pi}\Pi\sigma^{\mu\nu},\nonumber
\end{align}
with $D \equiv u^\mu\partial_\mu$, $\nabla^\mu \equiv \Delta^{\mu\nu}\partial_\nu$ and $\sigma^{\mu\nu} \equiv \nabla^{\langle\mu}u^{\nu\rangle}$, with $\langle\cdot\rangle$ denoting symmetrization and removal of the trace.

Here holography implies that after the fast hydrodynamization first order hydrodynamics is accurate, which implies $\pi^{\mu\nu} = 2\eta\sigma^{\mu\nu}$ and $\Pi = -\zeta\nabla\cdot u$\@.
We now construct an initial stage which uses a parameter $r_\text{hyd}$ to interpolate between the original free streaming pre-hydrodynamic stage and one inspired by the holographic computation done in \cite{1307.2539}\@. Such a construction interpolates between weak and strong coupling, and by varying $r_\text{hyd}$ in our Bayesian analysis we can attempt to answer the question as to whether experimental data prefer a weakly or a strongly coupled pre-hydrodynamic stage.

The explicit construction first computes the stress tensor at time $\tau = \tau_\text{hyd}$ using free streaming in the same way as described before, and decomposes it according to Eq.~\eqref{eq:decomp}\@. As $\rho$ and $u^\mu$ are the same for both free streaming and holography up to $\mathcal{O}(\tau_\text{hyd}^2)$, they are kept \footnote{An early implementation interpolated $\rho$ and $u^\mu$ between free streaming and holography as well, using the full holographic result in Eq.~\ref{eq:preflow}\@. This causes numerical issues when $\mathcal{T}$ is small, so the version described in the main text was used instead.}, and $\Pi$ and $\pi^{\mu\nu}$ are saved as $\Pi_\text{fs}$ and $\pi_\text{fs}^{\mu\nu}$, respectively. New values for $\Pi$ and $\pi^{\mu\nu}$ are then computed by using $r_\text{hyd}$ to interpolate between the free streaming result and the holographic result as follows:
\begin{align*}
\pi^{\mu\nu} & = r_\text{hyd}\pi_\text{hyd}^{\mu\nu} + (1 - r_\text{hyd})\pi_\text{fs}^{\mu\nu}, \\
\Pi & = r_\text{hyd}\Pi_\text{hyd} + (1 - r_\text{hyd})\Pi_\text{fs},
\end{align*}
with $\pi_\text{hyd}^{\mu\nu}$ and $\Pi_\text{hyd}$ the hydrodynamized shear and bulk term, respectively. In this way, $r_\text{hyd} = 0$ corresponds to weak coupling (free streaming) whereas $r_\text{hyd} = 1$ corresponds to strong coupling (holography inspired)\@. The result of the interpolation is then reinserted into Eq.~\eqref{eq:decomp}, and the result is used as the input for hydrodynamics.

From the discussion above, one might expect that $\pi_\text{hyd}^{\mu\nu}$ and $\Pi_\text{hyd}$ are given by $2\eta\sigma^{\mu\nu}$ and $-\zeta\nabla\cdot u$, respectively. In a Bayesian analysis however, it turns out this naive choice causes problems for certain rare combinations of parameters. In particular, for choices of $\eta/s$ which increase sharply with temperature in combination with small $\tau_\text{hyd}$ and large norm $N$, the amount of viscous entropy production is so large that the resulting final state contains an enormous amount of particles incompatible with observations. This large particle number however causes problems in the afterburner and the emulator, preventing the Bayesian analysis from rejecting such combinations of parameters. To solve this issue, one needs to take the second order terms into account, which depend quadratically on derivatives of $u^\mu$, and hence bring such extreme configurations back to reasonable values for particle yields, which are high enough to be excluded by the Bayesian analysis, but not high enough to cause problems in the computation. To include second order terms, however, we need to be careful what we mean by hydrodynamization of $\pi^{\mu\nu}$ and $\Pi$\@. The definition we take is %
that the hydrodynamized shear and bulk terms are what they would relax to given enough time, i.e.~assuming that $\tau_\pi$ and $\tau_\Pi$ are negligibly small. This in practice corresponds to ignoring the terms on the left-hand sides of Eqs.~\eqref{eq:bulk} and \eqref{eq:shear} and solving for $\pi^{\mu\nu}$ and $\Pi$, e.g.
\begin{align}
-\Pi & \equiv  \zeta\nabla\cdot u + \delta_{\Pi\Pi}\nabla\cdot u\Pi - \lambda_{\Pi\pi}\pi^{\mu\nu}\sigma_{\mu\nu},\nonumber \\
-\pi^{\mu\nu}& \equiv - 2\eta\sigma^{\mu\nu}\label{eq:shearhyd}  + \delta_{\pi\pi}\pi^{\mu\nu}\nabla\cdot u - \phi_7\pi_\alpha^{\langle\mu}\pi^{\nu\rangle\alpha}\nonumber \\
& \quad + \tau_{\pi\pi}\pi_\alpha^{\langle\mu}\sigma^{\nu\rangle\alpha} - \lambda_{\pi\Pi}\Pi\sigma^{\mu\nu},\nonumber
\end{align}
The resulting $\pi^{\mu\nu}$ and $\Pi$ are then saved as $\pi_\text{hyd}^{\mu\nu}$ and $\Pi_\text{hyd}$, respectively.
In practice, this can be easily done, as except for the $\phi_7$-term every term is linear in $\pi^{\mu\nu}$ and $\Pi$\@. The $\phi_7$-term itself turns out to be a small correction, which can be included by using the Newton-Raphson method starting from an ansatz where $\phi_7 = 0$\@. It is important that this definition is not much different from the naive first order definition if second order gradients are small, which is usually the case and a prerequisite for hydrodynamization. By construction, however, this definition guarantees that the bulk and shear pressure do not evolve rapidly at the time \tauhyd{}\@.

An important point is that $\pi_\text{hyd}^{\mu\nu}$ and $\Pi_\text{hyd}$ are qualitatively different from $\pi_\text{fs}^{\mu\nu}$ and $\Pi_\text{fs}$\@. In particular, in \cite{2010.15134}, it was shown that while $\Pi_\text{fs}$ is mostly positive and large, $\Pi_\text{hyd}$ is much smaller and negative. As $\Pi$ has a large effect on radial flow, this causes measurable differences in observables. If one uses a free streaming velocity different from the speed of light, \cite{2010.15130} shows that one can obtain similar profiles to $\Pi_\text{hyd}$, and in \cite{2010.15130,2010.15134} this free streaming velocity was used as a proxy for a strongly coupled pre-hydrodynamic stage in a Bayesian analysis. This resulted in a free streaming velocity compatible with a small value for $\Pi$ at the start of hydrodynamics. In this work, we improve on this simple scheme by actually interpolating between free streaming and strong coupling. In principle this can be improved further, perhaps replacing this scheme by attractor-like behavior \cite{1908.02866}\@.

\subsection{Posterior distribution for \rhyd}

\begin{figure}[ht]
\includegraphics[width=0.7\columnwidth]{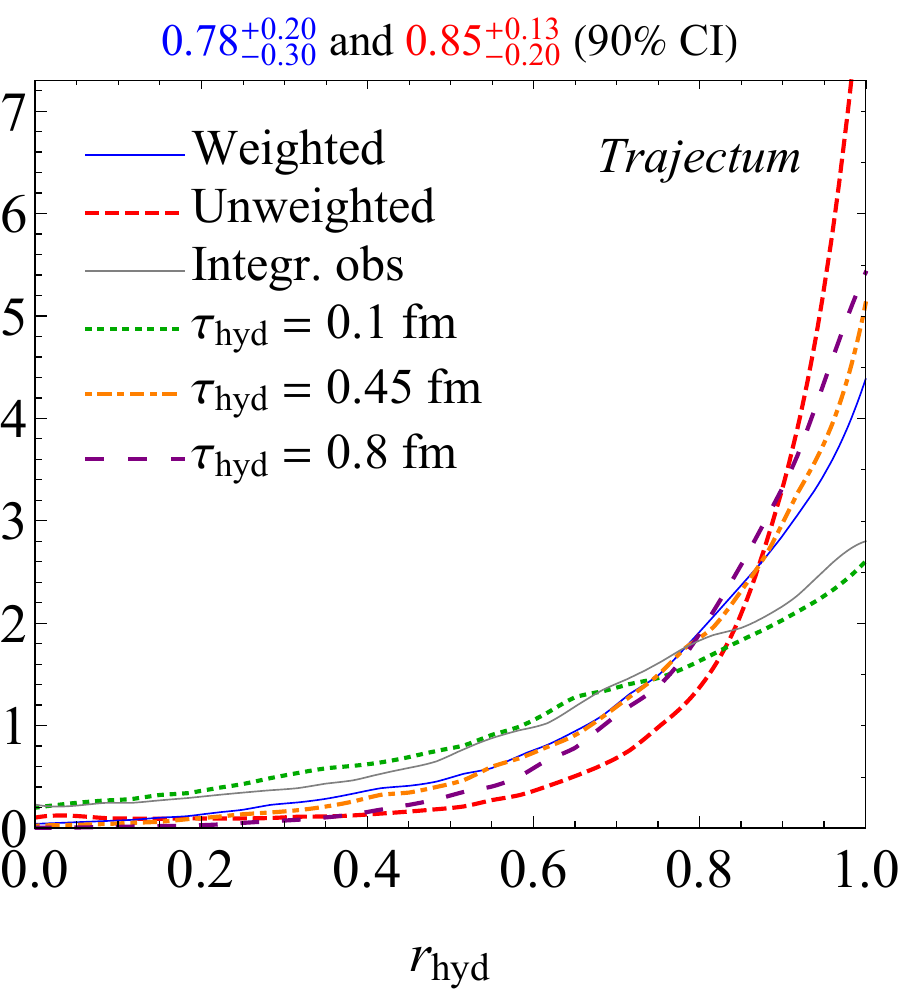}
\caption{\label{fig:rhydroposterior}We show the posterior for the \rhyd{} parameter that interpolates between free streaming ($r_\text{hyd}=0$) and a holographic far-from-equilibrium stage ($r_\text{hyd}=1$)\@. All settings show significant preference for the holographic scenario, which is strongest for unweighted observables or for late hydrodynamization times.}
\end{figure}

\begin{figure}[ht]
\includegraphics[width=0.5\columnwidth]{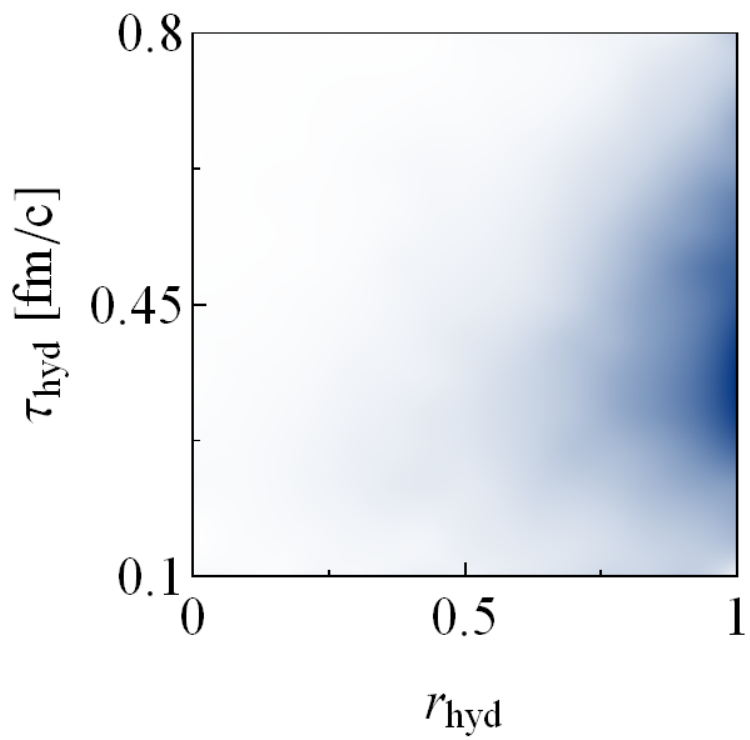}
\caption{\label{fig:rhydtauhydcorr}We show the correlation between \rhyd{} and \tauhyd{}\@. As in Fig.~\ref{fig:rhydroposterior} the posterior prefers the holographic scenario $r_\text{hyd}=1$\@. It is interesting that in that scenario \tauhyd{} is less constrained, which indicates an approximate independence of \tauhyd{} as expected in a hydrodynamized scenario.}%
\end{figure}

Fig.~\ref{fig:rhydroposterior} shows the posterior for \rhyd, again for the weighted and unweighted fitting scenarios. The posterior is strongly peaked at $r_\text{hyd} = 1$, 
indicating that our implementation for the strongly coupled pre-hydrodynamic stage is more compatible with experimental data than the free streaming weakly coupled one. We also included results that use only integrated observables as opposed to our full set including $p_T$-differential observables (gray)\@. This set also favors the strongly coupled scenario, albeit at lower significance. This shows that even with a lower weight for this parameter $p_T$-differential observables can make an important difference.
The intermediate values of \rhyd{} are harder to interpret, as they correspond to a linear interpolation between the two schemes. In all scenarios $r_\text{hyd} = 1$ is preferred over any intermediate value, so that the optimal value is physically well motivated. 

We also added three scenarios where \tauhyd{} equals 0.1, 0.45 or $0.8\,$fm$/c$ respectively. Clearly all three again favor the strongly coupled approach, but we note that $\tau_\text{hyd} = 0.1\,$fm$/c$ is significantly less sensitive to \rhyd{}\@. One reason may be that at such an early time the plasma has not hydrodynamized and e.g.~second and higher order gradients are still important. This would prohibit the use of the holographic equations as an input for hydrodynamics and may explain why the strongly coupled scenario is less favored. A second effect is that since free streaming takes the fluid away from hydrodynamization, a longer free streaming time makes the preference for a hydrodynamized fluid at the starting time of hydrodynamics more pronounced. Shortly we will find that $\tau_\text{hyd} = 0.4\,$fm$/c$ is preferred, but that $\tau_\text{hyd}$ is not strongly constrained.

In Fig.~\ref{fig:rhydtauhydcorr} we show the correlation of the posterior between \rhyd{} and \tauhyd{}\@. It is interesting that for the holographic scenario of $r_\text{hyd}=1$ the hydro starting time is significantly less constrained. This is a key signature of a smooth transition from a far-from-equilibrium stage to the hydrodynamic stage \cite{1307.2539}\@. Indeed after the fast hydrodynamization the holographic scenario should work well at any time until the approximation in Eq.~\eqref{eq:preflow} is not valid anymore.

An alternative analysis could have performed two separate runs for $r_\text{hyd}=0$ and $1$ and compared the corresponding Bayes factors as in \cite{2010.03928}, but such an approach would almost double the computational time if performed with two separate runs and does not necessarily add information beyond the current results since the relative Bayes factors are simply the ratio of the posterior at $r_\text{hyd} = 0$ and $r_\text{hyd} = 1$\@.

\section{The \emph{Trajectum} framework and the MCMC setup}\label{sec:trajectum}

\subsection{The initial state and hydrodynamics}

In this section we outline the \emph{Trajectum} 1.3 framework \cite{trajectumcode} and the Markov Chain Monte Carlo (MCMC) setup, in particular highlighting differences with respect to previous implementations \cite{2010.15134, 2110.13153} other than the already presented generalizations of the initial state and initial stage.
\emph{Trajectum} implements the initial state, pre-hydrodynamic stage, hydrodynamic stage, and freeze-out procedure of a heavy ion collision. \emph{Trajectum} is capable of producing 3+1D simulations, but in this work we assume boost invariance. %

For the initial state each Pb nucleus is composed of nucleons distributed according to a Woods-Saxon distribution \cite{Woods:1954zz}:
\[
\rho(r,\theta) \propto \frac{1}{1 + \exp\left(\frac{r - R}{\sigma}\right)},
\]
where $R$ and $\sigma$ are parameters. As in \cite{2112.13771}, we use different values for the Woods-Saxon parameters for protons and neutrons, simulating the effect of a neutron skin. In particular, we use $R_p = 6.68\,\text{fm}$, $R_n = 6.69\,\text{fm}$, $\sigma_p = 0.447\,\text{fm}$ and $\sigma_n = 0.56\,\text{fm}$ \cite{1710.07098}\@. Each nucleon then contains $n_c$ constituents, which are Gaussian sources of the thickness functions $\mathcal{T}_{A/B}$ of width $v$, distributed in a Gaussian distribution with width such that the whole nucleon has width $w$\@. Note here that $n_c$ can be non-integer, in which case one of the neighboring integers is chosen as the number of constituents as in \cite{2110.13153}, so that observables end up being continuous in $n_c$\@. In practice, $v$ is not directly used as a parameter, but we rather use $\chi_\text{struct}$ as follows \cite{1808.02106}:
\[
v = v_\text{min} + \chi_\text{struct}(w - v_\text{min}),
\]
with $v_\text{min} = 0.2\,\text{fm}$\@.

For each pair of nucleons, where one is taken from nucleus $A$ and the other from nucleus $B$, it is first determined whether the two nucleons collide as in \cite{1412.4708,1804.06469,1808.02106}\@. If they do, both nucleons are marked as \emph{wounded}\@. The constituents of each wounded nucleon in nucleus $A$ ($B$) then source the thickness function $\mathcal{T}_A$ ($\mathcal{T}_B$) as Gaussian sources of width $v$\@. The norm of each Gaussian source is given by $\gamma/n_c$, where for each constituent an individual $\gamma$ is sampled from a gamma distribution with mean 1 and width $\sigma_\text{fluct}\sqrt{n_c}$, with $\sigma_\text{fluct}$ yet another parameter which controls how much the energy deposition from each nucleon can fluctuate. The two thickness functions $\mathcal{T}_A$ and $\mathcal{T}_B$ are subsequently combined, and the result is used in the pre-hydrodynamic stage. The way in which the thickness functions are combined is modified in this work with respect to previous works, which is one of our main results already presented in %
Sec.~\ref{sec:q}\@. The initial state is followed by a pre-hydrodynamic stage which interpolates between weakly and strongly coupled schemes, which is also discussed already in detail in Sec.~\ref{sec:rhydro}\@. 

The pre-hydrodynamic stage initializes the stress tensor $T^{\mu\nu}$ at a time $\tau_\text{hyd}$, after which the stress tensor is evolved according to second order hydrodynamics in the 14 moment approximation \cite{1403.0962}, with the hybrid equation of state combining hadron resonance gas (HRG) and lattice QCD given by \cite{1804.06469,0912.2541,1407.6387}\@. %
The equations of motion are shown already in Eqs.~(\ref{eq:bulk}--\ref{eq:shear}) %
and here we only list the transport coefficients contained within them. The first order transport coefficients $\eta$ and $\zeta$ have a non-trivial temperature dependence, which we parameterize for the specific shear viscosity $\eta/s$ and the specific bulk viscosity $\zeta/s$, with $s$ the entropy density. The specific shear viscosity $\eta/s$ is given by a linear interpolation between the following points
\begin{align*}
\eta/s|_{T = 0\,\text{GeV}} & = \overline{\eta/s} - 0.075\cdot(\eta/s)_\text{slope}, \\
\eta/s|_{T = 0.15\,\text{GeV}} & = \overline{\eta/s} - 0.075\cdot(\eta/s)_\text{slope}, \\
\eta/s|_{T = 0.3\,\text{GeV}} & = \overline{\eta/s} + 0.075\cdot(\eta/s)_\text{slope}, \\
\eta/s|_{T = 0.5\,\text{GeV}} & = \overline{\eta/s} + 0.275\cdot(\eta/s)_\text{slope} + 0.2\cdot(\eta/s)_{\delta\text{slope}}, \\
\eta/s|_{T = 0.8\,\text{GeV}} & = (\eta/s)_{0.8\,\text{GeV}},
\end{align*}
where $\eta/s$ above $0.8\,\text{GeV}$ is constant, and $\overline{\eta/s}$, $(\eta/s)_\text{slope}$, $(\eta/s)_{\delta\text{slope}}$ and $(\eta/s)_{0.8\,\text{GeV}}$ are parameters, where $(\eta/s)_\text{slope}$ and $(\eta/s)_{\delta\text{slope}}$ are in units of $\text{GeV}^{-1}$, while the other two parameters are dimensionless. If $\eta/s$ is negative we replace it by $\eta/s = 0$\@.

This parametrization of $\eta/s$ differs in important aspects from previous implementations, where $\eta/s$ was parametrized by some constant value and one or several slopes \cite{Bernhard:2019bmu, 2010.15134, 2110.13153, 2011.01430})\@. In particular, physically it is expected that the QGP is mostly sensitive to the average, or effective, shear viscosity \cite{2010.11919} in approximately the 150 to 300\,MeV temperature region. The new parametrization with average and slope can be linearly mapped to the old constant plus slope and hence the Jacobian does not change the prior probability distribution. It can however be expected that we can get narrower constraints on $\overline{\eta/s}$ than on the previous $(\eta/s)_{\rm min}$ constant variable, which means we can focus on a smaller prior parameter range and hence improve precision in that region. Moreover, in the posterior distributions we get more direct constraints on the average $\eta/s$ as well as the slope.

By introducing the $(\eta/s)_{\delta\text{slope}}$ parameter it is possible to have a different slope above a temperature of 300\,MeV\@. From \cite{2010.11919} it is expected that heavy ion collisions do not have much sensitivity to this high temperature region, and this is indeed what we will find. Nevertheless, since $\eta/s$ enters in our initial stage model of pre-hydrodynamic strongly coupled evolution (see Sec.~\ref{sec:rhydro}) it is important that $\eta/s$ is still reasonable at the relatively high temperatures encountered at the early hydro switching time \tauhyd{}\@.

The specific bulk viscosity $\zeta/s$ is given by
\[
\zeta/s(T) = \frac{(\zeta/s)_\text{max}}{1 + \left(\frac{T - (\zeta/s)_{T_0}}{(\zeta/s)_\text{width}}\right)^2},
\]
with $(\zeta/s)_\text{max}$, $(\zeta/s)_\text{width}$ and $(\zeta/s)_{T_0}$ parameters. Earlier Bayesian analyses \cite{1808.02106,Bernhard:2019bmu} show that $(\zeta/s)_\text{max}$ and $(\zeta/s)_\text{width}$ are strongly negatively correlated. This is due to the fact that the actual simulations are mostly sensitive to the time averaged bulk viscosity, which, due to the gradual cooling down of the fluid, is proportional to $(\zeta/s)_{\text{m}\times\text{w}} \equiv (\zeta/s)_\text{max} \times (\zeta/s)_\text{width}$\@. For this reason, in our Bayesian analysis, we replace $(\zeta/s)_\text{width}$ by $(\zeta/s)_{\text{m}\times\text{w}}$ as the parameter that we vary. As we will show in Sec.~\ref{sec:allposteriors}, $(\zeta/s)_\text{max}$ and $(\zeta/s)_{\text{m}\times\text{w}}$ are indeed uncorrelated. Note that this implicitly changes our priors through the Jacobian associated with this transformation.

There are also second order transport coefficients, given by the shear relaxation time $\tau_\pi$, the bulk relaxation time $\tau_\Pi$, as well as $\delta_{\pi\pi}$, $\phi_7$, $\tau_{\pi\pi}$, $\lambda_{\pi\Pi}$, $\delta_{\Pi\Pi}$ and $\lambda_{\Pi\pi}$\@. The corresponding parameters are the dimensionless ratios
\[
\frac{\tau_\pi sT}{\eta}, \quad \frac{\tau_\Pi sT(1/3 - c_s^2)^2}{\zeta}, \quad \frac{\tau_{\pi\pi}}{\tau_\pi}, \quad \frac{\delta_{\pi\pi}}{\tau_\pi},
\]
\[
\phi_7P, \quad \frac{\lambda_{\pi\Pi}}{\tau_\pi}, \frac{\delta_{\Pi\Pi}}{\tau_\Pi}, \quad \frac{\lambda_{\Pi\pi}}{\tau_\Pi(1/3 - c_s^2)},
\]
with $c_s$ the speed of sound and $P$ the pressure. Of these, we vary the first and third in the Bayesian analysis, while we keep the others fixed to the kinetic theory values found in \cite{1403.0962}\@.

After hydrodynamics, the fluid is frozen out according to the Cooper-Frye procedure \cite{Cooper:1974mv} with the Pratt-Torrieri-Bernhard (PTB) prescription for viscous corrections \cite{1003.0413,1804.06469}\@. This occurs at the switching temperature $T_\text{switch}$, which is varied as a parameter in this work. From each event, we sample not one but five sets of particles, thereby improving statistics. 

The last parameter which is varied is a recent feature of SMASH \cite{Weil:2016zrk,dmytro_oliinychenko_2020_3742965,Sjostrand:2007gs}, which scales all interaction cross sections by a factor $f_\text{SMASH}$\@.

\subsection{Experimental data and Bayes theorem}
The experimental dataset used in this work is almost the same as in \cite{2110.13153}, with the addition of hadronic cross section measurements for PbPb and $p$Pb. We also added integrated unidentified anisotropic flow coefficients in the 0--1\% centrality bin and identified particle spectra in the 0.25--0.5\,GeV transverse momentum bin as a function of centrality. This together makes for a total of 653 individual data points (all data points are displayed in the appendix of \cite{2206.13522})\@. The complete experimental dataset then consists of the aforementioned PbPb \cite{2204.10148manual} and $p$Pb \cite{1509.03893} cross sections, as well as charged particle yields at 2.76 \cite{1012.1657} and $5.02\,\text{TeV}$ \cite{1512.06104}\@. We also include identified particle yields $dN_\text{ch}/dy$ and mean transverse momentum $\langle p_T\rangle$ for pions, kaons and protons at $2.76\,\text{TeV}$ \cite{1303.0737}, as well as unidentified transverse energy $E_T$ \cite{1603.04775} and fluctuations of mean $p_T$ \cite{1407.5530} at $2.76\,\text{TeV}$\@. We include the integrated anisotropic flow coefficients $v_2\{2\}$, $v_2\{4\}$, $v_3\{2\}$ and $v_4\{2\}$ at both 2.76 and $5.02\,\text{TeV}$ \cite{1602.01119}\@. We also include $p_T$-differential observables with bin boundaries at $(0.25, 0.5, 0.75, 1.0, 1.4, 1.8, 2.2, 3.0)\,\text{GeV}$\@. In particular, this includes spectra for pions, kaons and protons at $2.76\,\text{TeV}$ \cite{1303.0737}, as well as $v_2\{2\}(p_T)$ for pions, kaons and protons, and $v_3\{2\}(p_T)$ for pions (these data are only available for $p_T > 0.5\,$GeV) \cite{1606.06057}\@.

As was further motivated in \cite{2206.13522}, we optionally weight observables (we will show results both with and without weighting)\@. We define a weight $\omega$ to mean that we multiply the difference in an observable between theory and experiment by $\omega$, which preserves the correlation matrix. In particular, we have three classes with weight different from unity, as all of these classes are difficult to model theoretically. If an observable is in more than one of these classes, the weights multiply. In particular, we assign weight $1/2$ to any particle identified observable, weight $(100 - c[\%])/50$ for any centrality class $c$ over 50\% and $(2.5 - p_T[\text{GeV}]/1.5$ for $p_T$-differential observables with $p_T > 1\,\text{GeV}$ (except for protons since thermal protons generically have relatively large $p_T$ as compared to lighter hadrons)\@.

Having specified the experimental dataset we can determine the posterior probabilities $\mathcal{P}(\boldsymbol{x}|\mathbf{y}_{\exp})$ of our parameters according to Bayes theorem
\begin{equation}
    \mathcal{P}(\boldsymbol{x}|\mathbf{y}_{\exp})
    = \frac{e^{-\Delta^2/2}}{\sqrt{(2\pi)^{n} \det\left(\Sigma(\boldsymbol{x})\right)}} \mathcal{P}(\boldsymbol{x}) 
    \label{eq:bayes}
\end{equation}
with $\mathcal{P}(\boldsymbol{x})$ the (flat) prior probability density and where
\begin{equation}
    \Delta^2
    = \left(\mathbf{y}(\boldsymbol{x})-\mathbf{y}_{\rm exp}\right)\cdot \Sigma(\boldsymbol{x})^{-1} \cdot \left(\mathbf{y}(\boldsymbol{x})-\mathbf{y}_{\rm exp}\right),
    \label{eq:delta}
\end{equation}
with $\mathbf{y}(\boldsymbol{x})$ the predicted data for parameters $\boldsymbol{x}$, $\mathbf{y}_{\rm exp}$ the $n$ experimental data points and $\Sigma(\mathbf{x})$ is the sum of the experimental and theoretical covariance matrices. The covariance matrices are constructed as in \cite{1804.06469}\@.

Much of the implementation of the Markov Chain Monte Carlo to determine the posterior distributions for our parameters has been explained elsewhere \cite{2010.15134,1804.06469,1904.08290} and here we hence only sketch the main idea and then focus on specific changes we made. The standard procedure for a Bayesian analysis as presented here is to determine model parameters at a relatively modest number of design points that are distributed on a latin hypercube within the 23-dimensional parameter space \footnote{We note that this includes the variation of $\snn$, but since we fix this to its respective values of 61.8 and 67.2\,mb at 2.76 and 5.02\,TeV respectively, it is not displayed elsewhere in this work.}\@. In this work we use 1200 design points, where at each point we constructed 60k initial conditions, of which we evolved 15k hydrodynamic evolutions (sampled quite evenly, but higher in the 0--1\% centrality bin)\@. From each hydro event we ran 5 SMASH events, whereby we only evolve particles created between pseudorapidity $-1.5$ and $1.5$\@. In this way the statistical uncertainties are subdominant for almost all data points used and we could furthermore also include the ultracentral 0--1\% anisotropic flow values as an observable.

With the resulting design run it is then possible to train Gaussian Process Emulators (using Sci-kit) on the Principal Components (PCs) of the observables of interest. Crucially the emulator can not only predict observables at any point in the 23-dimensional parameter space, but it also provides an estimate of its own uncertainty, including correlations between observables as encoded in the PC transformation. Given such a fast evaluation of the model it is possible to construct posterior distributions in Eq.~\eqref{eq:bayes} given the experimental data using the parallel tempered emcee code \cite{B509983H, 1202.3665}\@.

We now highlight a few differences in our implementation with respect to the previous versions of \cite{1804.06469, 2110.13153}\@.

\subsection{Observable classes}\label{sec:classes}
With the standard approach the emulator uncertainty for the $\saa$ cross section is still unnecessarily large. Without observable classes and for our settings the emulator estimates its own uncertainty at 3.14\%, which is in fact a significant overestimate from the 1.4\% coming from the explicit validation of the emulator. Still, both uncertainties are of the same order as the experimental uncertainty (also 3.1\%) and hence a reduced emulator uncertainty leads to improved constraints. Since $\saa$ only depends significantly on mainly the nucleon width (see \cite{2206.13522}) it is in fact straightforward to emulate with high precision. The reason why the emulator nevertheless is not so accurate lies in the PC transformation, which mixes much more difficult observables with $\saa$\@.

The two reasons for the PC transformation are firstly the computational speed and secondly to properly take into account correlations coming from the emulator uncertainty. The latter correlations, however, are relatively small among unlike observables, such as the spectra versus anisotropic flow observables. For this reason in this work we divided our observables in three classes:
\begin{itemize}
    \item The cross section
    \item Multiplicities, spectra and mean transverse momenta
    \item All anisotropic flow observables.
\end{itemize}
For all these classes we did separate PC analysis (25 PC for each, or fewer if there are fewer observables in the class)\@. This greatly improved the emulator uncertainty of $\saa$ down to 0.24\%, but also the average predicted emulator uncertainty for all other observables went down from 10.0\% to 7.7\%  due to more precise multiplicities and spectra. The classes are hence quite essential for precision for $\saa$, but also improve the accuracy in general.

\subsection{The covariance matrix at MAP}\label{sec:covatmap}

In the current analysis we do not compute the full emulator uncertainty at each evaluation in the MCMC algorithm. Instead, we evaluate the uncertainty matrix at the most likely parameter setting (Maximum A Posteriori, MAP), whereby we update this point every 100 steps. The prime reasoning for this change was an observed phenomenon where the MCMC would attribute a high likelihood to regions in phase space with a large emulator uncertainty instead of a good description of the data. From a Bayesian point of view this may seem sensible: since for some datapoints our model cannot find a fully satisfactory description (see also the appendix in \cite{2206.13522}) it will try to move the parameter space to regions where the model is less well understood. In some way this is the opposite from the `searching under a lamppost effect'\@. Evaluating the covariance at the MAP point avoids the MCMC to walk towards these regions that do not describe the data as well. Models that increase design points in `likely regions' will suffer even more from this effect \cite{2102.11337, 2302.14184}, and can hence likely also benefit from this new method.

One may wonder what would happen if by chance a MAP point would lie in these less accurate regions, after which a higher likelihood will be found due to the larger uncertainties in the covariance matrix. In this case, however, the next 100 steps will converge the posterior towards regions that describe the data better (leaving the covariance matrix fixed) and the new MAP point will hence lie again in the region that has the better description. 

It is an interesting question if this new evaluation method is fully consistent with Bayesian reasoning, but we think it provides at least a better physical description with realistic estimates of the emulator uncertainties. Of course if the posterior space is not too large we do not expect a large variation in the covariance matrix and indeed we will see that there is only an insignificant change in the posterior. The new method also has the advantage that it is at least ten times faster, making it feasible to run a full MCMC analysis in one day on a single computer.

We note that both this change and the change in the observable classes have no impact if the emulator uncertainty is significantly smaller than the experimental uncertainty. For many observables this is the case for experimental uncertainties at 2.76\,TeV (see \cite{2110.13153}), but especially for new precise $5.02\,$TeV data and for anisotropic flow coefficients the emulator uncertainty is dominant and hence important.

\begin{figure*}[ht]
\includegraphics[width=\textwidth]{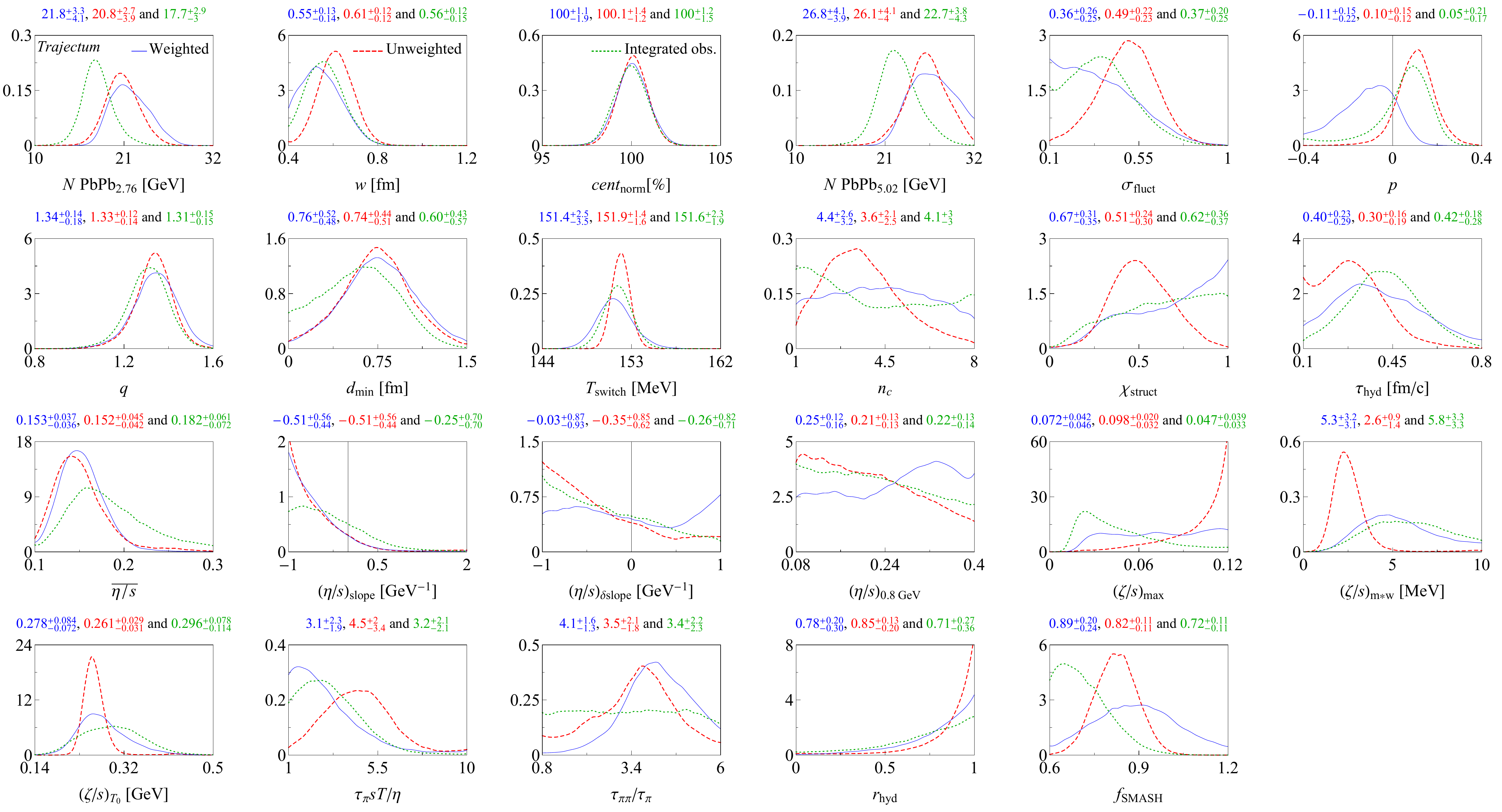}
\caption{\label{fig:allposteriors}%
We show the posterior for all our parameters for the weighted (blue), unweighted (red dashed) and unweighted integrated only observables (green dotted)\@. Most parameters are in agreement between the three scenarios, though we note that especially the bulk viscosity is much more constrained without weighting observables. The weighted scenario is hence more conservative but likely also more realistic. The $p_T$-differential observables (not included for the green curve) provide tighter constraints for mostly the shear viscosity.}
\end{figure*}

\subsection{The centrality normalization}\label{sec:centnorm}
As described in detail in \cite{2110.13153} (see also \cite{2204.10148manual, 1301.4361}) it is important to vary the centrality normalization since there is both a theoretical and experimental uncertainty on how many hadronic collisions occur given a (measured) luminosity which also contains an uncertainty. This uncertainty is especially important for observables that are highly centrality dependent, such as multiplicities and spectra in peripheral events, but even elliptic flow at central collisions depends significantly on the centrality normalization. Currently the experimental uncertainty is estimated at only 1\% \cite{2204.10148manual}, but we note that this translates to uncertainties of up to 10\% in multiplicity for the more peripheral events.
A major advantage of including this uncertainty as a parameter in a global analysis rather than including it as a systematic uncertainty is that it automatically includes the correlations among all observables. Indeed for this uncertainty for instance spectra and elliptic flow are highly correlated, which would be difficult to take into account in any other way.

In this work we introduced a prior distribution for the centrality normalization as a Gaussian with width 1\%, which is equal to the experimental uncertainty \cite{2204.10148manual}\@. This is somewhat exceptional with respect to the other parameters, where we have a flat prior distribution, but in this case there is a clear special reason why this prior knowledge is there. Nevertheless, shortly in the next subsection we will vary this prior to a flat prior and see that the difference is relatively mild.

It is an interesting question if we should have varied the centrality normalizations for 2.76 and 5.02\,TeV collisions separately. Indeed both data sets are significantly different, both regarding detector effects as well as multiplicity. Nevertheless, given the similarity in the method to obtain the centrality distribution the uncertainties are likely highly correlated and hence it is most sensible to assume the centrality normalization for both energies to be equal. 

\section{Posterior distributions}\label{sec:allposteriors}

\begin{figure*}[ht]
\includegraphics[width=\textwidth]{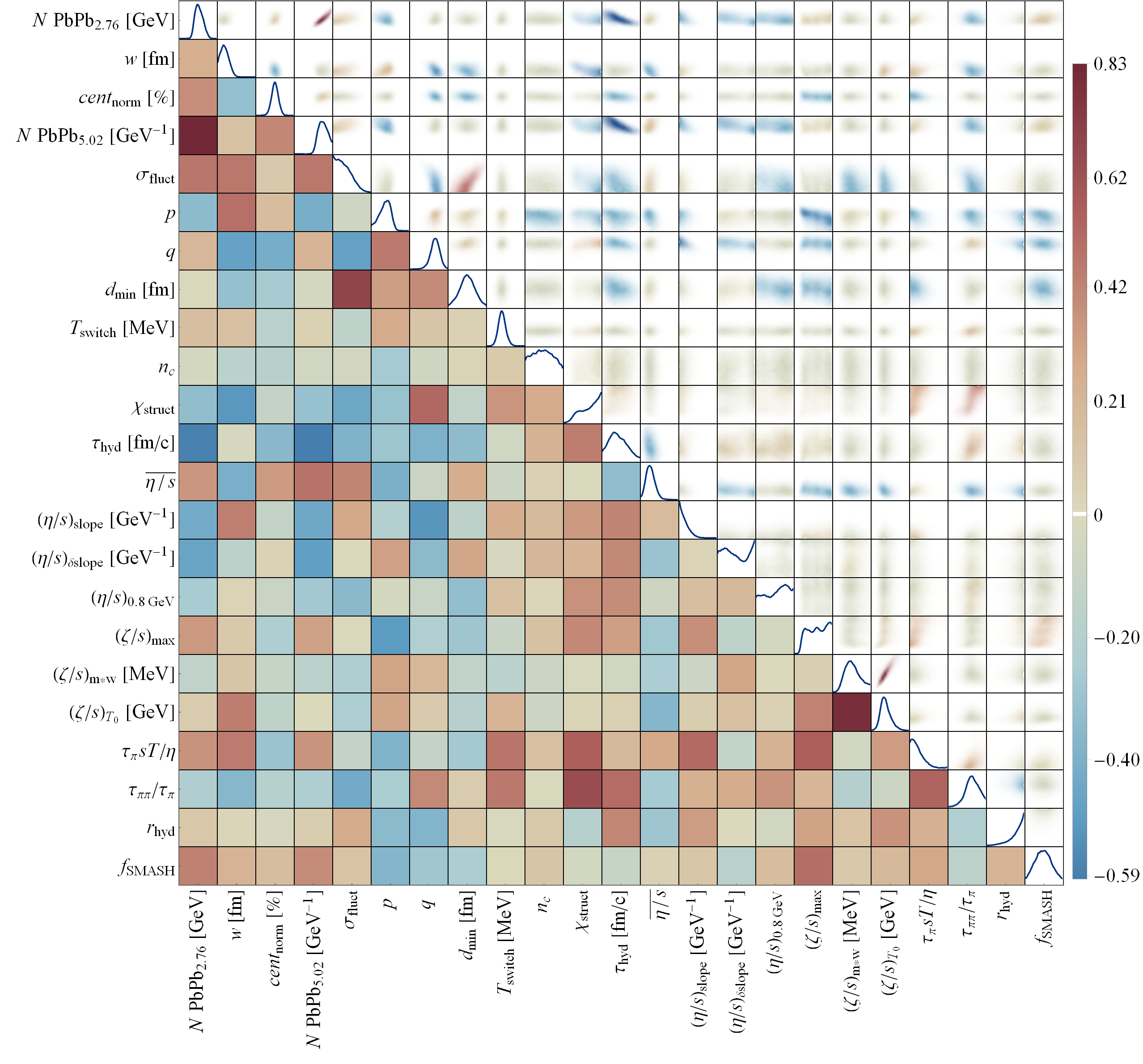}
\caption{\label{fig:correlations}We show the full correlation matrix for all our parameters. The color coding on the left triangle indicates the Pearson correlation of all 2D distributions shown in the right triangle. The ranges of the parameters equal those of Fig.~\ref{fig:allposteriors}\@.}
\end{figure*}

Fig.~\ref{fig:allposteriors} shows the posteriors for all parameters varied in the Bayesian analysis both with (blue, solid) and without (red, dashed) observable weighting \footnote{Though essentially the same as in \cite{2206.13522}, the simulations presented here have a significantly higher precision. In particular the best estimate for the nucleon width decreased from $0.62^{+0.18}_{-0.17}$ to $0.55^{+0.13}_{-0.14}\,$fm at 90\% confidence interval.}\@. We also included an analysis without $p_T$-differential observables (green, dotted), whereby we did not include weighting. It can be seen that in general the posterior distributions become broader with observable weighting as compared to without. This can be understood because observable weighting decreases the discrepancy between theory and experiment artificially, leading to less extreme Bayes factors. 

Weighting gives a higher weight to more robust observables, such as $\saa$\@. It is hence no surprise that this leads to a smaller nucleon width $w$, which is preferred by the relatively low $\saa$ \cite{2206.13522}\@. The unweighted distributions are sometimes much more peaked than the weighted ones, which is in particular the case for $\zetamax$ and $(\zeta/s)_{\text{m}\times\text{w}}$\@. The interpretation is that $p_T$-differential observables provide strong constraints on the bulk viscosity, but as explained we do not think this is fully physically trustable due to the theoretical uncertainties in especially high $p_T$ bins. The weights are perhaps somewhat arbitrary, but in this case provide a more conservative estimate on the uncertainty of the posterior distributions. Of course the posterior results are especially trustworthy if all three methods give consistent answers, which is the case for most parameters.

Interestingly the weighted distributions give stronger constraints on the second order transport coefficients $\tau_\pi s T/\eta$ and $\tau_{\pi\pi}/\tau_\pi$, even though they are consistent with the unweighted analysis. They are also consistent with \cite{2010.15130}, even though that analysis used a considerably different model. We will see shortly that the second order parameters are quite highly correlated with structure parameters, making this more non-trivial. The agreement is a good indication that indeed global Bayesian analysis can constrain second order transport coefficients.

\begin{figure*}[ht]
\includegraphics[width=\textwidth]{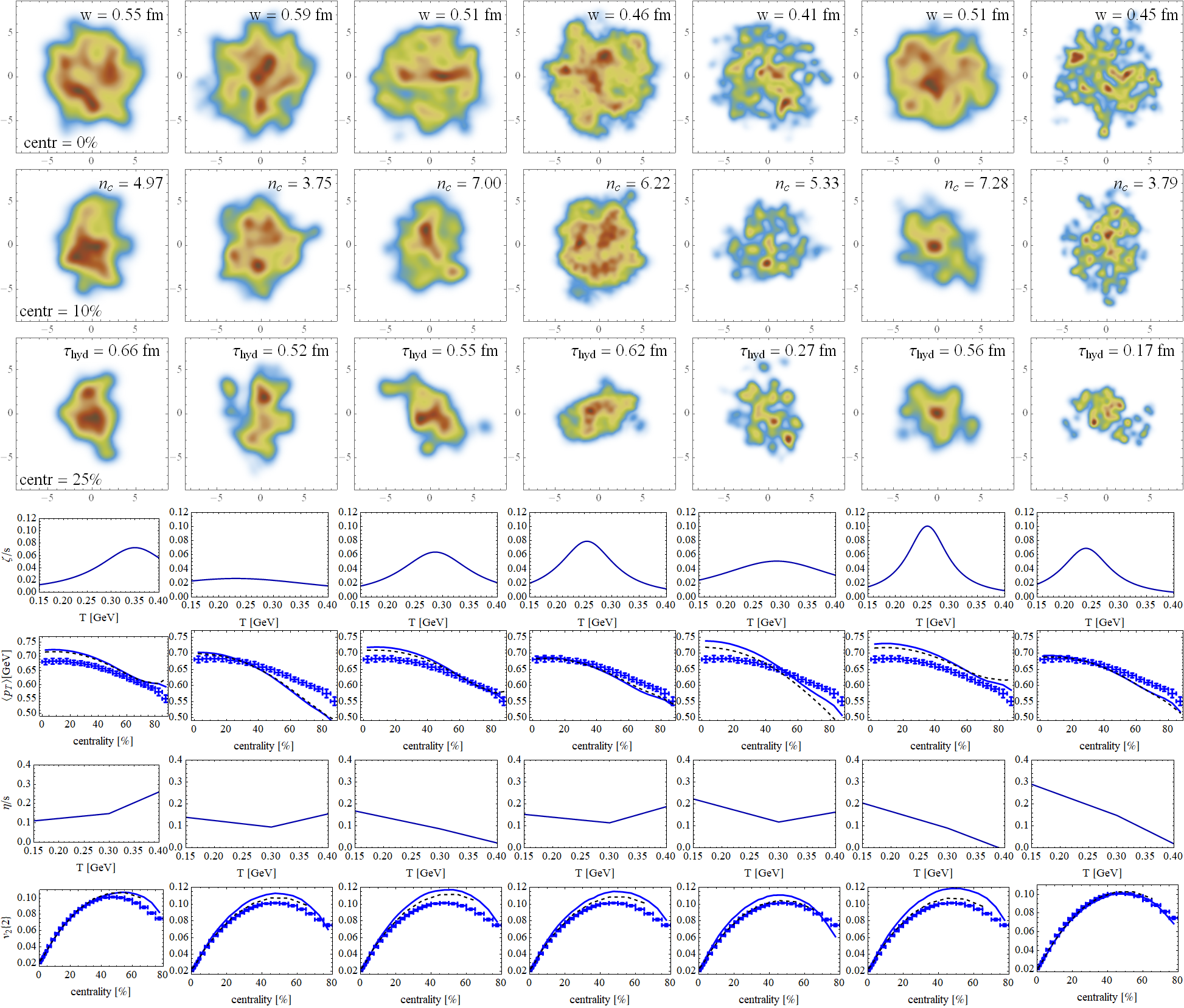}
\caption{\label{fig:PbPbsummary}For seven probable settings drawn from the MCMC we show the energy density at starting time of hydrodynamics $\tau_\text{hyd}$ for three centrality classes (top three rows)\@. The different settings have characteristically different specific shear and bulk viscosities ($\eta/s$ and $\zeta/s$) that lead to different values for the mean transverse momentum (\meanpt) or elliptic flow (\vtwo) of charged particles (bottom four rows)\@. The experimental data is compared both with the emulator (dashed) as well as an explicit computation (solid) at the corresponding parameter point.}
\end{figure*}

A parameter of note is the new SMASH parameter $f_\text{SMASH}$, which modifies the interaction probabilities inside SMASH by an overall factor. The posterior shows that $f_\text{SMASH}$ is consistent with unity, but not well constrained when using observable weighting, indicating that the overall interaction strength in SMASH has little effect on the observables we use. This lends credence to the idea that the strengths of particular individual interactions, many of which are not measured, are not very important for results such as ours, removing a source of modelling uncertainty.

Fig.~\ref{fig:correlations} shows the correlations between pairs of parameters for the posterior which includes observable weighting. In Sec.~\ref{sec:q}, it was noted that $E_\text{ref}$ was chosen such that $q$ is relatively uncorrelated with the norm $N$\@. One can see that indeed $q$ is not strongly correlated with the norm at either 2.76 or $5.02\,\text{TeV}$\@.

In general, the reason one prefers correlation to be small is that preferentially the posterior distribution should be well contained in the prior range, while simultaneously one wants the prior ranges to be as small as possible. The former is needed so that the prior does not artificially constrain the posterior, while the latter is required because smaller priors lead to smaller emulator uncertainties. If two parameters are strongly correlated, this means that a relatively large prior range must be chosen to contain the entire posterior distribution, where the posterior will be close to zero in a large part of such a prior range. A notable exception to this rule is the correlation between the norms at different energies. Since a single emulator is used for both energies this correlation does not increase the emulation uncertainty.

A good example of this (anti-)correlation can be found in \cite{Bernhard:2019bmu}, where the maximum of the joint posterior distribution for $(\zeta/s)_\text{max}$ and $(\zeta/s)_\text{width}$ occurs in an area shaped like a hyperbola, with tails of the distribution not being contained in the prior range. In Fig.~\ref{fig:correlations}, one can see that replacing $(\zeta/s)_\text{width}$ by $(\zeta/s)_{\text{m}\times\text{w}}$ indeed removes this anticorrelation. However, there is still a tail to the distribution which is not contained within the prior though, which is both the region with large  $(\zeta/s)_\text{max}$ as well as the region with large  $(\zeta/s)_{\text{m}\times\text{w}}$\@. The first is relatively easy to explain: given that we keep $(\zeta/s)_{\text{m}\times\text{w}}$ fixed, a large $(\zeta/s)_\text{max}$ effectively makes the bulk viscosity a delta distribution. The statement that large $(\zeta/s)_\text{max}$ is not excluded hence implies that currently the model cannot exclude a bulk viscosity that acts like a delta distribution. 

The tail of $(\zeta/s)_{\text{m}\times\text{w}}$ is more subtle. Firstly, we note that there is a strong correlation (0.80) with the temperature $(\zeta/s)_{T_0}$\@. This is relatively natural, since the evolutions spent less time at high temperature (where roughly $T\propto \tau^{-1/3}$) and hence a significant bulk viscosity has more influence at lower temperatures. We can conclude, however, that a large $(\zeta/s)_{T_0}$ is ruled out. This is perhaps surprising (see also \cite{Bernhard:2019bmu}), since a large $(\zeta/s)_{T_0}$ in combination with a large $(\zeta/s)_\text{width}$ would result in a more or less constant bulk viscosity. What this shows is that this is not consistent with experimental data, presumably since the model has a strong preference for a small bulk viscosity around the switching temperature (see also \cite{2206.13522})\@.
In future work, one could try to remove this correlation as well by replacing $(\zeta/s)_{T_0}$ by $(\zeta/s)_{T_0}/(\zeta/s)_{\text{m}\times\text{w}}$ as the varied parameter, which should be uncorrelated with $(\zeta/s)_{\text{m}\times\text{w}}$\@.

Several other interesting correlations can be seen in Fig.~\ref{fig:correlations}\@. The nucleon width is strongly anticorrelated (Pearson correlation $-0.43$) with the \trento{} $q$ parameter. This has a relation to the change in \trento{} describing an energy density or an entropy density. Indeed, the original Bayesian analysis in \cite{1605.03954} used \trento{} as an estimator for the initial entropy density (equivalent to approximately $q=4/3$) and found a small nucleon width of $0.48\,$fm, whereas later analyses used \trento{} as an energy density ($q=1$) and found a large nucleon width of approximtely 1\,fm. Still, as argued in \cite{2206.13522}, this is only partially the explanation for our small nucleon width, whereby the main effect is the inclusion of the total $\saa$ cross section. Indeed, when analyzing our chain restricting $q=1$ we still find a relatively small width of $w=0.67\pm0.14\,$fm. %

Another strong correlation is between \dmin{} and \sigmaf{} (0.54)\@. This is sensible: increasing \dmin{} spreads out nucleons more evenly along the nucleus and hence decreases fluctuations. In this analysis it is relatively new that we cannot rule out \sigmaf{} being zero. This is in contrast to \cite{Bernhard:2019bmu, 2010.15130} where \sigmaf{} was significantly positive. Here also the small nucleon width is important, as the width is positively correlated with \sigmaf{} since a smaller nucleon automatically makes a more fluctuating profile. Even with a small width and small fluctuations we still find a positive \dmin{}, which indicates that a nucleon within a nucleus has a short-distance repulsive core.

\begin{figure*}[ht]
\includegraphics[width=\textwidth]{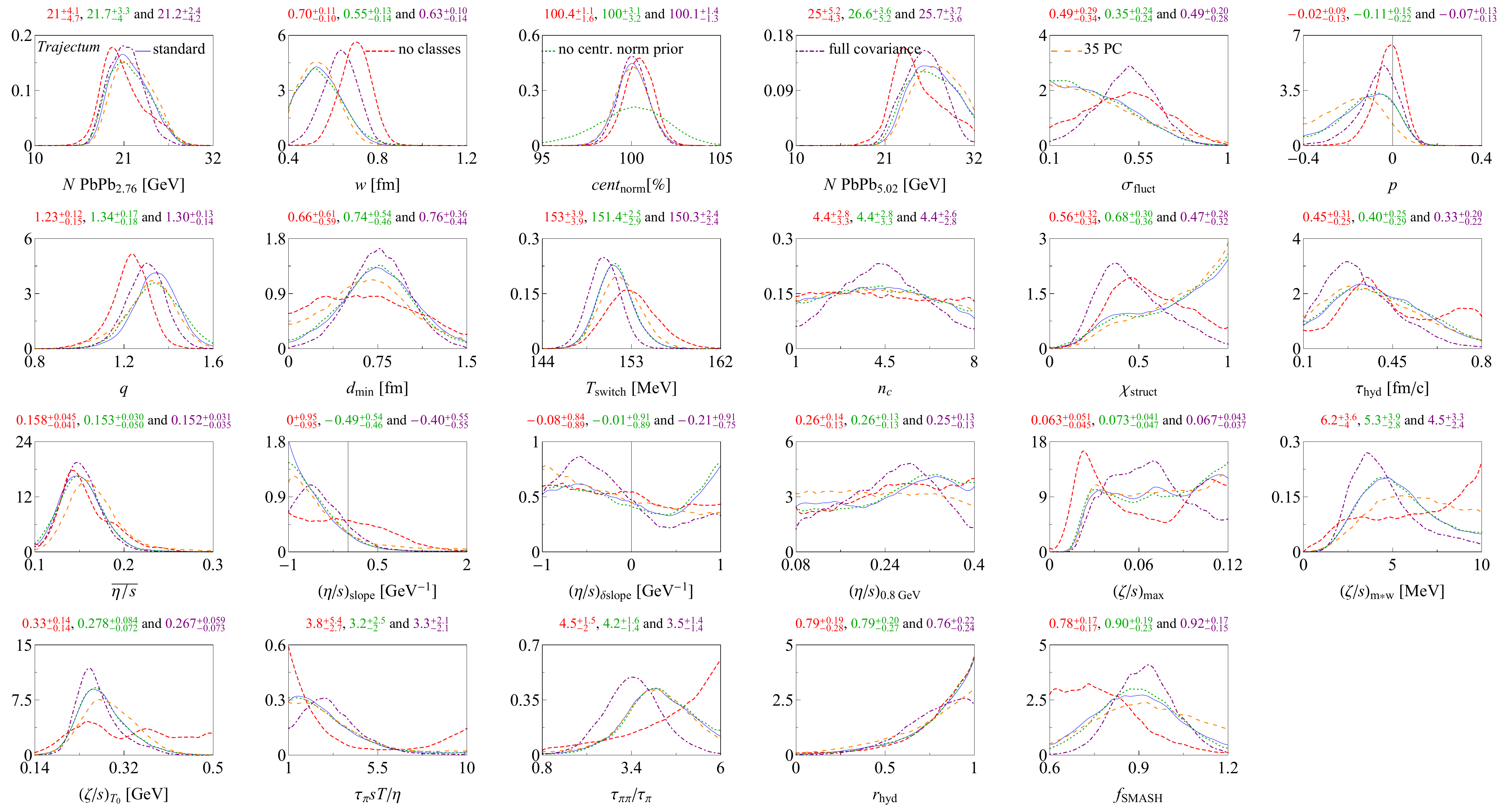}
\caption{\label{fig:allposteriorsvary}We show the same results as in Fig.~\ref{fig:allposteriors}, but including several variations in the MCMC procedure as described in the main text. While there are significant differences in the curves themselves overall they are consistent. Especially using observable classes improves the precision of estimating $\saa$ and thereby gives a better estimate of the nucleon width (see also \cite{2206.13522})\@.}
\end{figure*}

Many correlations exist between in particular substructure parameters (such as the subnucleon width \chistruct) and second order transport coefficients. Even with these correlations we managed to obtain mild constraints on these parameters. It is however important to note that for these constraints it is then essential to use a model with subnucleonic substructure. Without such substructure one would obtain artificially strong constraints on especially \taupipi{}\@.

Lastly, in Fig.~\ref{fig:PbPbsummary} we have randomly drawn seven example settings from the posterior and show the initial energy density profile for three centrality classes, the respective shear and bulk viscosities versus temperature and finally the comparison with data of the mean transverse momentum (\meanpt) and the elliptic flow (\vtwo) for both the emulator prediction (dashed) as well as a direct computation (solid)\@. It is interesting to see that even after a full global analysis fitting to 653 data points the initial energy density is still consistent with a wide range of profiles that differ visibly by eye. An important factor here is the time (early profiles are less smooth), but also other parameters are important. Smoother profiles lead to smaller viscosities, even though this effect is somewhat hard to discern by eye. 

What is slightly worrying is that the emulator fits to especially \vtwo{} are in much better agreement with experiment than the direct computation. We verified that the difference is fully consistent with the emulator uncertainty, which varies between 6\% and 9\%, but the consistent overestimate requires further study (our uncertainty is however comparable to \cite{1804.06469, 2302.14184})\@. Clearly for anisotropic flow observables the emulator uncertainty is much larger than the experimental uncertainty and increasing the number of design points would help significantly to constrain the model.

\subsection{MCMC variations}

To improve our understanding of the improvements in the MCMC algorithm and also to test its robustness we compare in Fig.~\ref{fig:allposteriorsvary} our weighted posterior (blue solid) with several variations. In the first (red, dashed) we repeated our analysis, but without using the observable classes as described in Sec.~\ref{sec:classes} (or equivalently, using a single observable class)\@. As explained this significantly increases the emulation uncertainty on $\saa$, whereby $\saa$ was the prime driver for a smaller nucleon width. It is hence not surprising that this setting gives a larger nucleon width ($w = 0.71$ versus 0.55 fm)\@. Overall all results are however still consistent.

In green dotted we show a variation with a flat prior probability density for the centrality normalization. Understandably this makes the posterior distribution significantly less constrained, but the most likely value is still approximately 100\%, indicating that the centrality normalizations within \emph{Trajectum} and the ALICE experiment are fully consistent. The prior also has very little effect on the other parameters, mostly since the normalization does not have strong correlations with the other parameters (see Fig.~\ref{fig:correlations})\@. The only mild correlation is with the $q$ parameter, such that indeed the $q$ posterior of this variation is wider than in the standard approach.

In purple dot-dashed we show the analysis when evaluating the covariance matrix at every step in the MCMC, which hence evaluates Eq.~\ref{eq:bayes} exactly. While as explained in Sec.~\ref{sec:covatmap} we think that evaluating the covariance matrix at the MAP point gives perhaps more physical results it is comforting to see that both approaches make very little difference. This is also expected when the emulator uncertainty is relatively small or constant. An extra advantage is the enormous speed-up of the standard implementation. Lastly, in orange we verified that the number of PCs used does not significantly affect our results.

\section{Discussion}

There are two main results in this work, corresponding to Sec.~\ref{sec:q} and \ref{sec:rhydro}, respectively. Bayesian analysis is naturally suited to discriminate between different microscopic physics scenarios, where as long as we can interpolate between different such scenarios with parameters this can be done relatively cheaply in terms of computation time. The T\raisebox{-0.5ex}{R}ENTo model was introduced exactly as such an interpolation between different qualitative features of microscopic initial state models. It was able to provide evidence for EKRT-like scaling, while providing evidence against wounded nucleon and KLN scaling. %
In this work, we have extended the T\raisebox{-0.5ex}{R}ENTo formula so that we can also reproduce binary scaling behavior as well as the scaling behavior exhibited by for example IP-Glasma, and found that binary scaling is strongly disfavored, while the scaling behavior of IP-Glasma is compatible with our findings. 

The same extension allows us to ask the question whether to interpret the result of the T\raisebox{-0.5ex}{R}ENTo formula as an energy density or an entropy density. Interestingly, one would be led to believe that it should be interpreted as an energy density, until one forces compatibility with the PbPb hadronic cross section as in \cite{2206.13522}, which favors interpretation as an entropy density. As was also pointed out in \cite{2208.06839}, this puts the favored initial state back to the first Bayesian analysis from the Duke group \cite{1605.03954}, with nucleons around $0.4\,\text{fm}$ and the result of the T\raisebox{-0.5ex}{R}ENTo formula once again being interpreted as an entropy density. Here we note that the model in \cite{1605.03954} is simpler, and that the reasons for getting smaller nucleons were mostly the lack of radial flow in the pre-hydrodynamic stage. The re-interpretation of the result of the T\raisebox{-0.5ex}{R}ENTo formula as an energy density without a parameter such as $q$ to allow the Bayesian analysis to correct for this change in interpretation in \cite{Bernhard:2019bmu} caused this question to be unaddressed until now.

One way in which this analysis is radically different from all previous analyses, however, is the pre-hydrodynamic stage. Where in earlier analyses free streaming was used, which essentially assumes zero coupling, in this analysis we use a parameter, $r_\text{hyd}$, to interpolate between free streaming (zero coupling) and a scheme modelled after holography (infinite coupling)\@. Allowing the Bayesian analysis to find an optimal value for $r_\text{hyd}$, we find that the holographic scheme is strongly preferred over free streaming, where we take note that this preference is present irrespective of whether one uses observable weighting or not. 
The preference for holography may also be present for $\tau_\pi s T /\eta$, which equals  $4-\log(4)\approx 2.61$, \cite{Baier:2007ix} in holography and $5$ \cite{Denicol:2014vaa} at weak coupling (see also \cite{2010.15130})\@. For $\tau_{\pi\pi}$ both the holographic ($88/35(2-\log(2))\approx 1.92$, \cite{0809.4272}) and the weak coupling ($10/7 \approx 1.43$, \cite{Molnar:2013lta}) values are unlikely according to this analysis.

\begin{table}[ht]
\begin{tabular}{cccccc}
\hline
\hline
& Std. & $r_{\text{hyd}} = 0$&$q = 1$&$f_{\text{SMASH}} = 1$ \\

\hline
 $dN_\text{ch}/d\eta$ & 0.67 & 0.61 & 0.57 & 0.70 \\
 $dN_{\pi^\pm,k^\pm,p^\pm}/dy$ &  0.98 & 0.98 & 0.87 & 1.00 \\
 $dE_T/d\eta$ & 2.19 & 2.04 & 1.85 & 2.29 \\
 $\langle p_T\rangle_{\text{ch},\pi^\pm,K^\pm,p^\pm}$ &  0.89 & 0.91 & 0.75 & 0.93 \\
 $\delta p_T/\langle p_T\rangle$ & 0.65 & 0.64 & 0.58 & 0.62 \\
 $v_n\{k\}$ &  0.61 & 0.58 & 0.63 & 0.61 \\
 $d^2N_{\pi^\pm}/dy\,dp_T$ & 1.45 & 1.34 & 1.23 & 1.50 \\
 $d^2N_{K^\pm}/dy\,dp_T$ & 1.79 & 1.69 & 1.55 & 1.84 \\
 $d^2N_{p^\pm}/dy\,dp_T$ &  1.67 & 1.65 & 1.47 & 1.74 \\
 $v_2^{\pi^\pm}(p_T)$ & 0.96 & 0.70 & 1.19 & 1.04 \\
 $v_2^{K^\pm}(p_T)$ & 1.12 & 0.94 & 1.31 & 1.11 \\
 $v_2^{p^\pm}(p_T)$ & 0.55 & 0.53 & 0.60 & 0.49 \\
 $v_3^{\pi^\pm}(p_T)$ & 0.68 & 0.73 & 0.51 & 0.62 \\
$\saa$ &1.13 & 1.23 & 1.83 & 1.18  \\
\hline
 \text{average} & 1.10 & 1.04 & 1.07 & 1.12 \text{} \\
\hline
 \end{tabular}
\caption{Average number of standard deviations from experimental data for different classes of observables for our standard fit together with three fixed values of the new parameters introduced in this work per observable class when used in the weighted analysis. Uncertainties include experimental uncertainty and theoretical uncertainty from the emulation (the latter is dominant for the $v_n$ classes)\@. Surprisingly the average deviation for $r_{\rm hyd} = 0$ is slightly lower than for the standard analysis even though $r_{\rm hyd} = 0$ is ruled out in the full posterior when including all correlations. \label{tab:observables}}
\end{table}

An important question is what in the experimental data points to this holographic model as opposed to free streaming. In Tab.~\ref{tab:observables} we see that this is not a straightforward question. In fact, most observables get slightly better if $r_{\rm hyd} = 0$, which is also reflected in the average deviation going down from 1.10 to 1.04\@. In reality the Bayes factor is more complicated, since the Bayes factor is the sum of the squares of the differences in standard deviations, whereas we show the sum of the absolute values. Also, some observable classes contain more datapoints and also the correlations between the datapoints are very important. For $q = 1$ it is clearer that this is not preferred since otherwise it is difficult to properly describe $\saa{}$\@. For completeness we also added one more column with our last new parameter $f_\text{SMASH}$ fixed to 1\@. Given that the standard approach is consistent with $f_{\text{SMASH}} = 1$ it is unsurprising that this leads to similar results.

An interesting future avenue of research would be to further explore the pre-hydrodynamic stage. In our AdS/CFT-inspired scheme we have simply assumed that the timescales $\tau_\pi$ and $\tau_\Pi$ governing the relaxation to hydrodynamics are much smaller than other timescales, but in reality this assumption is probably not warranted, as due to the rapid longitudinal expansion both $\rho$ and $u^\mu$ are not really constant at the time $\tau_\text{hyd}$\@. In the future, one could construct a more sophisticated pre-hydrodynamic stage which takes this into account, perhaps taking inspiration from attractor-like behavior \cite{1908.02866, 1805.01604}\@. One could subsequently use a parameter similar in spirit to $r_\text{hyd}$ to interpolate between this improved scheme and ours, thereby using Bayesian analysis to test the merits of such an improved approach.

Another possible improvement is to try to further reparameterize the transport coefficients which we take to have a non-trivial temperature dependence, $\eta/s$ and $\zeta/s$\@. This should ideally result in a parameterization which is broad enough to cover a wide range of physics scenarios, while allowing for the posterior distribution to be efficiently contained within the prior range. A related problem is the fact that some currently allowed parameter combinations result in transport coefficients whose non-trivial features fall outside the range probed by the simulations.

Finally, it would be interesting to extend this work beyond boost invariance. Bayesian analyses in 3+1D have been performed before \cite{1610.08490}, albeit with a simpler pre-hydrodynamic stage. In principle it is straightforward to perform the required 3+1D simulations, and indeed the \emph{Trajectum} code is capable of doing so. However, the computational cost of such simulations is an order of magnitude larger than the 2+1D simulations used here, so in practice this would require careful planning to manage computational cost.

As with all Bayesian analyses we should end here with a word of caution. Even within our model it is clear that adding new elements in the model (such as second order transport) affects the posterior distributions of the other parameters. This is sensible, since a larger model space gives the global analysis more space to obtain a proper description of the data. Even though the current model is likely the most versatile for soft physics currently on the market it is clear that many physical elements are still missing and of potential importance. One example mentioned is relaxing boost invariance and an even more sophisticated initial stage, but also the particlization procedure is somewhat limited (see for instance \cite{2010.03928}) and at the moment we do not have propagating non-hydrodynamic degrees of freedom (either because gradients are large \cite{1404.7327} or due to thermal fluctuations \cite{2208.04806})\@. 

A second standard limitation of Bayesian analyses is the fact that relatively small differences in many data points (or their correlations) add up to large Bayes factors that can artificially constrain the posterior distribution. This happened for instance in older analyses of the nucleon width, which turned out inconsistent with new data (the total hadronic cross section $\saa{}$)\@. A relatively crude attempt to ameliorate this problem is to underweight (sets of) data points, which reduces this problem. Still, in an ideal world all uncertainties would be accurately estimated, including their full correlation matrix, which is a challenge for future studies.

\section*{Acknowledgements}
We thank Giuliano Giacalone, Scott Moreland and Krishna Rajagopal for interesting discussions. We thank Scott Moreland for sharing the output of his code at \cite{scott}\@.
GN is supported by the U.S.~Department of Energy, Office of Science, Office of Nuclear Physics under grant Contract Number DE-SC0011090.

\bibliographystyle{apsrev4-1}
\bibliography{prc, prcmanual}

\end{document}